\documentclass[12pt,preprint]{aastex}

\slugcomment{28 SEP 2012}
\shorttitle{ATCA Observations of Wd1}
\shortauthors{Fok et al.}

\begin{document}

\title{Maser Observations of Westerlund 1 and Comprehensive Considerations on Maser Properties of Red Supergiants Associated with Massive Clusters}

\author{Thomas K.T. Fok, Jun-ichi Nakashima, Bosco H. K. Yung, Chih-Hao Hsia}
\affil{Department of Physics, University of Hong Kong, Pokfulam Road, Hong Kong, China}
\email{junichi@hku.hk}

\and

\author{Shuji Deguchi}
\affil{Nobeyama Radio Observatory, National Astronomical Observatory of Japan, Minamimaki, Minamisaku, Nagano 384--1305, Japan}



\begin{abstract}
We report the results of Australia Telescope Compact Array (ATCA) observations of the Westerlund~1 (Wd1) region in the SiO $v=1$, $J=1$--$0$ and H$_2$O 6$_{16}$--5$_{23}$ maser lines, and we also report the analysis of maser properties of red supergiants (RSGs) associated with 6  massive clusters including Wd1. The primary purpose of this research is to explore possibilities of using maser emission for investigating the nature of massive clusters and associated RSGs. The SiO $v=1$, $J=1$--$0$ and H$_2$O 6$_{16}$--5$_{23}$ maser lines are detected toward 2 of 4 known RSGs in Wd1. The large velocity ranges of maser emission are consistent with the RSG status. RSGs with maser emission tend to exhibit redder $\log (F_{21}/F_{12})$ and [K--12.13] colors compared to RSGs with no maser emission. The mass-loss rates derived from dust radiative transfer modeling suggest that RSGs with maser emission tend to exhibit larger mass-loss rates compared to RSGs with no maser emission. In an extended sample of 57 RSGs in 6 massive clusters, detections in the SiO line tend to homogeneously distribute in absolute luminosity $L$, whereas those in the H$_2$O line tend to distribute in a region with large $L$ values.
\end{abstract}




\section{Introduction}

Until now, about 200 red supergiants (RSGs) have been identified in the Galaxy \citep{lev09}, but a majority of known RSGs are isolated field stars, which pose difficulties in obtaining critical stellar parameters, such as distance, age and luminosity. However, in the last several years some massive clusters harboring multiple RSGs have been found, including Westerlund 1 \citep[Wd1;][]{men07}, Red Supergiant Cluster 1 \citep[RSGC1;][]{fig06}, RSGC2 \citep{dav07}, RSGC3 \citep{ale09,cla09}, RSGC4 \citep{neg10a}, and RSGC5 \citep{neg11}. These clusters that harbor multiple RSGs at a time are suitable sites for investigating the nature of RSGs, because RSGs in a single cluster share a coeval condition and the same distance.

It is notable that RSGs often exhibit maser emission of the SiO, H$_2$O and OH molecules. Since at the frequencies of these maser lines, interstellar extinction is negligible, maser emission potentially could be a useful tool for investigating RSGs hidden in the galactic plane. However, maser emission of RSGs embedded in massive clusters still has not been well-studied. In order to explore possibilities of using maser emission for investigating massive clusters and embedded RSGs, we have made maser searches toward 5 massive clusters, which are observable from the northern hemisphere, in the SiO ($v=1$ and 2, $J=1$--0) and H$_2$O (22~GHz) maser lines \citep{nak06,deg10}, and until now roughly a dozen of maser sources of RSGs have been found in the 5 massive stellar clusters. In addition, the results of an SiO maser survey of Galactic RSGs in the SiO $v=1$, $J=2$--1 line has been recently reported \citep{ver12}.

In this paper, we report the results of interferometric maser observations using the Australia Telescope Compact Array (ATCA) toward the massive cluster Wd1 in the SiO and H$_2$O maser lines at 43 GHz and 22 GHz, respectively. The purposes of this research are (1) to search SiO and H$_2$O maser emission toward the Wd1 region and (2) to explore possibilities of using maser emission for investigating the nature of massive clusters and associated RSGs. To clarify the maser properties of RSGs in massive clusters, we analyze the previous data of maser observations of massive clusters, in conjunction with the present observation of Wd1.

The structure of this paper is as follows. In Section 2, we describe the details of the observations of Wd1. In Section 3, we summarize the results of SiO and H$_2$O maser observations of Wd1. In Section 4, we discuss the velocity information derived by the present observations. In section 5, we discuss the infrared properties of RSGs in Wd1 and other massive clusters. Finally, the main results of the present research are summarized in Section 6.


\section{Observations and Data Reductions}

The radio interferometric observation in the SiO $v=1$, $J=1$--$0$ line towards the Wd1 region was made with ATCA on 2008 November 20 and 21 (project code: C1796, PI: Nakashima, J.). Since the angular size of the Wd1 region \citep[$2' \times 4'$;][]{bor70} is much larger than that of a single field-of-view (FOV) of an ATCA 22~m antenna at the observing frequency of 43~GHz (HPBW$\sim66''$), we observed 8 FOVs with different center positions to cover all RSGs and RSG candidates in Wd1 (center positions of observed FOVs are summarized in Table~1). In Wd1, \citet{men07} have identified 4 RSGs (W~20, W~75, W~26 and W~237) through their near infrared spectroscopy, and all the 4 known RSGs were covered in the present observation (coordinate values of the known RSGs are summarized in Table~2). In addition to 4 known RSGs, to make doubly sure, a majority of $K$-band bright stars in the Wd1 region (i.e., possible mass-losing evolved stars) are covered with 8 FOVs. In Figure~1, we present the locations of 8 observed FOVs (see, the blue circles in Figure~1) and 4 known RSGs. 

The weather condition was stable throughout the entire observation run. The observation used the EW352/367 array configuration consisting of six 22~m antennas. The baseline lengths ranged from 50 to 4400~m. The rest frequency of the SiO $v=1$, $J=1$--$0$ line (43.122090~GHz) was centered at a correlator window with a bandwidth of 32~MHz. The frequency coverage was ranged from 43.148 to 43.116~GHz. The 256 channel correlator gave the effective velocity coverage and resolution of 222.6 km~s$^{-1}$ and 0.869 km~s$^{-1}$, respectively. The quasar 1253$-$055 (3C279) was observed at the beginning of the observation for bandpass and flux calibrations. A nearby strong SiO maser source, IRAS 16105$-$4205 was also observed at the beginning for checking reliability of the system \citep[the SiO $v=1$, $J=1$--$0$ line is detected for the first time toward this object, even though the strong $v=1$, $J=2$--$1$ line was previously detected; ][; see, Appendix A, for details]{hai94}. The resulting uncertainty of the absolute flux density is roughly within 15--20$\%$. Observations were made in a snap-shot mode, which observed 8 FOVs in turn. The observation was periodically interleaved with the nearby gain calibrators, 1646$-$50, to track the phase variations over time (the phase calibrator was observed for 3 minutes every roughly 30 minutes). After all positions were observed, the array slewed to the first position to loop again. At least 3 loops were repeated for each position throughout the whole observing session. The on-source integration time at each FOV was 10 minutes in each single loop except for positions 2 and 3, which took 15 minutes each (this is because positions 2 and 3 are the most crowded region in Wd1, a better signal-to-noise ratio helps to avoid confusion in source identification). Total integration time on position 1, positions 2--3, and positions 4--8 were 0.7, 1.0, and 0.5 hours, respectively. And another 1.6 hours were used for calibrations. Visibility data were calibrated using MIRIAD \citep{sau95}, following a standard calibration procedure. Then we applied a self-calibration technique using the MIRIAD task, {\it selfcal} to enhance the signal-to-noise ratio. Image processing of the data was also performed with MIRIAD by applying robust weighting, which compromises between the natural and uniform weighting (we applied ``robust=0.5"). As we observed different FOVs in a snap-shot mode, the $u$--$v$ coverages are different from FOV to FOV. Therefore, synthesized beams at each FOV have slightly different sizes and position angles; the information about synthesized beam sizes is summarized in Table~1 (note: averaged synthesized beam size is $6.5''\times2.4''$). We checked the detection of the continuum emission by integrating over all emission-free channels (the integration range is about 25~MHz), but did not detect it. The upper limit is $1.5\times10^{-3}$~Jy~beam$^{-1}$.

Archival ATCA data of a previous observation in the H$_2$O 6$_{16}$--5$_{23}$ line toward Wd1 (project code: C1619, PI : Sean Dougherty\footnote{The OH 1612~MHz line was also observed in this project, and the emission seems to be detected toward Wd1 according to our brief inspection. However, as the data quality is not good enough for further analysis, we do not use it in the present research.}) were used for the present analysis. The H$_2$O maser observation was made from 2006 December 6 to December 10 using the 6A array configuration. The six antennas form baselines ranged from 340 to 5900~m. The observed H$_2$O line at 22.2351204~GHz is centered at a correlator window with a bandwidth of 16~MHz. The frequency coverage is ranged from 22.243~GHz to 22.227~GHz. The 512 channel correlator gave an effective velocity coverage and resolution of 215.7~km~s$^{-1}$ and 0.421 km~s$^{-1}$, respectively. The HPBW of a single antenna is 126$''$ at the observing frequency. With this beam size of a single antenna, almost the entire cluster region was covered only with 2 FOVs. We present the locations of observed FOVs in Figure~1 (see, the green circles) and Table~1. The observation was made in a snap-shot mode as well as the SiO maser observation. A nearby phase calibrator, 1646$-$50 was observed every 10 minutes. A flux calibrator used was 1253$-$055 (3C279). Calibration and imaging of the data were performed using MIRIAD following the same methodology as adopted for the SiO data. We applied robust weighting (robust=0.5) giving an averaged synthesized beam size of $1.3''\times0.4''$. We did not detect any continuum emission (integration range is roughly 10~MHz); the upper limit is $2.4\times10^{-3}$~Jy~beam$^{-1}$.


\section{Results of Maser Observations of Westerlund 1}

\subsection{Sources Identification}
First of all, we briefly describe how we identified stellar maser sources in the data cubes of the present observation. Source identification processes are common in both SiO and H$_{2}$O observations. We first sorted out all emission features detected above a 3~$\sigma$ level in each velocity channel of calibrated data cubes, and obtained the coordinate values of each emission feature by fitting a two-dimensional Gaussian function using a MIRIAD task {\it imfit}. At this stage, roughly 10 emission sources are selected as possible detections in each data set of the SiO and H$_{2}$O observations. Then, we excluded artificial features in the following procedure.

Generally speaking, true maser lines exhibit a linewidth larger than, at least, 1--2~km~s$^{-1}$, which is equivalent to 2 and 3 channels at 43~GHz and 22~GHz, respectively (velocity resolutions of the present observation at 43 GHz and 22 GHz are 0.869~km~s$^{-1}$ and 0.421~km~s$^{-1}$, respectively, as we mentioned in Section 2). Therefore, we looked for the emission features that consistently appear at the same position within uncertainty of a synthesized beam size and survive respectively in more than 2 and 3 consecutive velocity channels at 43~GHz and 22~GHz, so that we could exclude spiky features with a very narrow linewidth, which are most likely artificial. 

Then, we also doubly checked the reliability of detections by cross-checking with 2MASS images. Since the photosphere of RSGs exhibits a low effective temperature of about 3000~K, a bright near IR counterpart with a red IR color should be found at the positions of maser sources if emission features are really originated in a RSG. We cross-checked between our radio images and 2MASS $K$-band images to find IR counterparts. As the angular resolution of 2MASS images is about 2.0$''$ in $K$-band \citep{skr06}, the resolution of the present observation seems to be sufficient enough for positional comparison. In considerations of beam sizes of both 2MASS and the present observations, we defined that emission is real if the 2MASS point sources with a red color are lying within 4.5$''$  and 2.0$''$ with respect to the radio positions of emission features at 43~GHz and 22~GHz, respectively. Through these steps, finally we identified two real emission sources detected each in the SiO and H$_2$O lines (the SiO and H$_2$O lines are detected toward the same two sources). The details of the results are summarized in Section 3.2.

\subsection{Results of the SiO and H$_2$O Maser Observations}

As mentioned in Section 3.1, we finally detected the SiO $v=1$, $J=1$--$0$ line toward 2 of 4 known RSGs (W~26 and W~237) in Wd1, and also detected the H$_2$O 6$_{16}$--5$_{23}$ line toward the same two RSGs. In addition, we inspected across entire observing regions to doubly check further maser detections of unknown mass-losing evolved stars, but no other detections were found at a 3~$\sigma$ level. In Table~3 we present the coordinate values of detected maser sources, peak and velocity integrated intensities, intensity peak velocities and rms noise levels. In Table~4 we present rms noise levels toward two known RSGs with no maser emission. In Figure~2 we present the line profiles of detected maser lines. 

Both W~26 and W~237 exhibit asymmetric profiles (asymmetric with respect to the peak velocity) with a wide velocity range; the velocity ranges of W~26 and W~237 are roughly 40 km~s$^{-1}$ and 23 km~s$^{-1}$, respectively. Generally speaking, RSGs exhibit a wider velocity range than that of asymptotic giant branch (AGB) stars, which exhibit a velocity range of, at most, 10 km~s$^{-1}$ in the SiO maser lines \citep[see, e.g.,][]{nak07,zha12a}. Therefore, the velocity range of SiO maser emission is consistent with the RSG status. Additionally, these two stars clearly exhibit asymmetric profiles (note: the black arrows in Figure~2 indicate weak component above a 5~$\sigma$ level). This is also typical characteristics often seen in RSGs \citep[see, e.g.,][]{nak07,zha12a}. Both W~26 and W~237 exhibit clearly different intensities in the SiO line (the peak intensities of W~26 and W~237 are 0.12~Jy and 1.54~Jy, respectively), while both stars exhibit almost the same intensity in the  H$_2$O line (the peak intensities of  W~26 and W~237 are 0.73~Jy and 0.72~Jy, respectively). Since the two stars are at the same distance as in the same cluster, the difference in intensity directly reflects the difference of the absolute intensity of maser emission. This difference in intensity in the SiO maser line may be explained by the time variation in intensity of maser emission \citep[see, e.g.,][]{kam05,zha12a}.

We shall note that the intensity of the H$_2$O maser line is somewhat weaker than that of three well-known nearby (isolated) RSGs: i.e., VY CMa (1.2~kpc), VX Sgr (1.6~kpc) and NML Cyg (1.6~kpc) exhibit the peak intensities of 199.8~Jy, 167.2~Jy and 18.8~Jy \citep[these flux values are scaled at 3.5~kpc, ][]{pas06,choi08,zha12b}, while W~237 and W~26 exhibit 0.72~Jy and 0.73~Jy. Here, we find roughly two order of differences in intensity between the two cases. Similarly, the SiO maser emission ($v=1$, $J=1$--0) of the three well-known nearby RSGs exhibit systematically larger fluxes than RSGs in Wd1: VY CMa, VX Sgr and NML Cyg exhibit the peak intensities of 5.4~Jy, 31.4~Jy and 2.6~Jy \citep[scaled at 3.5~kpc,][]{bob04,zha12a,su12}, while W~237 and W~26 exhibit 1.54~Jy and 0.12~Jy, respectively. However, these differences in intensity between nearby RSGs and RSGs associated with Wd1 are partly due to the uncertainty of distances. In fact, the three representative nearby RSGs exhibit relatively large 12~$\mu$m fluxes as well as the large maser intensities: i.e., the 12~$\mu$m fluxes of VY CMa, VX Sgr and NML Cyg are 1,166~Jy (IRAS), 433~Jy (MSX) and 587~Jy (MSX; these flux values are scaled at 3.5~kpc), while those of W~237 and W~26 are 101~Jy, and 326~Jy, respectively. The systematically large mid-infrared fluxes of the nearby RSGs means that the distances used for scaling may include a large uncertainty (even though a large dispersion in infrared intensities should be also taken into account). Although the distances to the nearby RSGs are based on the trigonometric parallax measurements giving a reliable distances, as we discuss later in Section 4, the distance to Wd1 possibly includes a large uncertainty. Therefore, for more precise discussions on absolute maser intensities, we need more reliable distance information about Wd1.

On the other hand, the intensity ratios of the H$_2$O to SiO maser intensities are distance-free parameters, and the nearby RSGs seem to exhibit a relatively large value on average: the peak intensity ratios of the H$_2$O to SiO maser lines of  VY CMa, VX Sgr and NML Cyg are 36.9, 5.3 and 7.3, while those of  W~237 and W~26 are 0.47 and 6.1. Even though the results is not very obvious (because only VY CMa and W~237 exhibit extreme values), one may consider that this relatively weak intensity of the SiO maser line of RSGs in Wd1 is somehow originated in the cluster environment; for example, the outer layer of the envelope may be disturbed by interactions with ambient materials in the condensed cluster environment.

It is notable that SiO maser emission of W~26 and W~237 exhibits larger velocity ranges than H$_2$O emission. In the case of spherical envelopes, the velocity range of SiO emission theoretically should be smaller than that of H$_2$O emission, because molecular gas is usually still being accelerated in the radial direction in the innermost region (2--3 R$_{\ast}$), from which SiO maser is emitted. Since maser emission cannot be amplified in the direction of acceleration, SiO masers are usually amplified only in the tangential direction to the radial direction. Therefore, in the case of a spherical envelope, intensity peaks of SiO maser emission are found at velocities close to the systemic velocity. On the contrary, at the H$_2$O maser region (20--30 R$_{\ast}$) the motion of molecular gases usually already reaches to almost the terminal velocity and, in addition, gas is usually turbulent. Therefore, the maser can be amplified in any direction. As a result, the velocity range of H$_2$O maser emission is generally wider than that of the SiO maser emission. The present result possibly suggests that circumstellar envelopes of observed RSGs somewhat deviates from spherical symmetry. We shall note, however, that this discussion may be affected by the time variation of the line profile and sensitivity of the observation if weak components exist below the detection limit (for example, as mentioned later in Section 4, the convective motion could alter the line profile).

\section{Systemic Velocities of Maser Sources}

In the case of AGB stars, it is well known that the peak velocity of SiO maser emission represents the stellar systemic velocity within an uncertainty of 2--3 km~s$^{-1}$ \citep{jew84}. In the case of RSGs, however, the peak intensity clearly does not represent the systemic velocity of the stars, because the line profiles are not symmetric in velocity. For that reason, here we estimate the systemic velocities of RSGs in Wd1 by assuming that the median velocity of a line profile is close to the stellar systemic velocity: specifically, we calculated the average of the highest and lowest velocities of 5~$\sigma$ channels. In the case of SiO maser emission, the derived values of the median velocities for W~26 and W~237 are $-38.1$~km~s$^{-1}$ and $-41.7$~km~s$^{-1}$, respectively. Similarly, the median velocities obtained from the H$_2$O maser profiles are $-39.3$~km~s$^{-1}$ and $-48.5$~km~s$^{-1}$ for W~26 and W~237, respectively (since W~26 exhibits a single-peak profile in the H$_2$O line, here we simply give the peak velocity). These derived values are not exactly consistent with the systemic velocities obtained by infrared high-dispersion spectroscopy of the CO band-head at 2.29~$\mu$m \citep[near-infrared velocities are $-44.5$~km~s$^{-1}$ and $-57.5$~km~s$^{-1}$ for W~26 and W~237, respectively;][]{men09}. The differences between the present observation and near-infrared measurement are roughly 5--6~km~s$^{-1}$ and 9--16~km~s$^{-1}$ for W~26 and W~237, respectively. 

As we mentioned in Section 2, we checked the reliability of our velocity calibration by observing a known maser source with a known systemic velocity, and confirmed that the present velocity measurement is consistent with a previous measurement \citep{hai94}. On the other hand, the resolution of near-infrared observations \citep{men09} was also pretty high ($R \sim 9000$; this corresponds to $\Delta v=33$~km~s$^{-1}$), and the accuracy of the infrared velocity measurement must reach to, at least, a several km~s$^{-1}$ level, because a series of absorption lines were simultaneously observed and the derived velocities were averaged out. Therefore, inconsistency in the velocities seems to be originated in astrophysical reasons. In fact, the atmosphere of RSGs are strongly affected by convective motions: the maximum expansion velocities of RSG envelopes could be often beyond 30--40~km~s$^{-1}$ due to convective motions \citep{jos05}. In addition, the infrared velocity determined by observing the CO band-head may be affected by the pulsation of RSGs; for example, in the case of M-type miras, the pulsation is known to shift the position of the CO band-head up to 15 km~s$^{-1}$ \citep{sch00}. Presumably, the difference in velocities up to 16~km~s$^{-1}$ may be interpreted for these reasons.

The average of two systemic velocities of SiO maser emission is calculated to be $-$39.9~km~s$^{-1}$. Similarly, in the case of H$_2$O maser emission, the average is $-$43.9~km~s$^{-1}$. These values are expected to be close to the systemic velocity of Wd1. Previously, two indirect measurements of the systemic velocity of Wd1 have been made based on the motion of ambient interstellar gas (HI and CO) surrounding Wd1:  $-$55$^{+9}_{-26}$~km~s$^{-1}$ \citep[HI;][]{kot07} and  $-48\pm14$~km~s$^{-1}$ \citep[CO;][]{lun09}. The present values are not inconsistent with these values if we consider the uncertainty (the systemic velocities of Wd1 are briefly summarized in Table~5).

Using the systemic velocity derived, we calculated the kinematic distance to Wd1 under the assumption of a flat circular rotation model of the Galaxy (here, we use the velocity of SiO maser emission as a representative value): in the calculation we followed the same methodology adopted by \citet{kot07}, assuming the galactocentric distance of 7.6 kpc, and the derived kinematic distance is $3.0\pm0.5$~kpc (here we assume an uncertainty of 10~km~s$^{-1}$ in the velocity of Wd1 based on the above discussion). We also calculated the kinematic distance based on \citet{rei09} for comparison, and found no significant differences from those based on \citet{kot07}: the obtained kinematic distances are summarized in Table~6. On the other hand, the obtained kinematic distance is slightly nearer than luminosity distances derived by independent methodologies using main-sequence stars, OB supergiants and Wolf-Rayet stars in Wd1 \citep[3.5--5.5 kpc,][]{cro06,bra08,neg10b}. Therefore, in the following analysis,  we adopt 3.5 kpc as the representative distance to Wd1, because this distance is not inconsistent with both the kinematic and luminosity distances [at the distance of 3.5~kpc, absolute luminosities of W~26 and W~237 are respectively calculated to be $1.1 \times 10^{6}$ L$_{\odot}$ and $2.3 \times 10^{5}$ L$_{\odot}$ based on the photometric data mentioned in Section 5.1; these values are roughly consistent with a range of typical absolute luminosities of RSGs \citep[$2 \times 10^{4}$ L$_{\odot}$ --- $6 \times 10^{6}$ L$_{\odot}$; see, e.g.,][]{sto99,mau11}] .


\section{Discussion}
Our primary purpose of this project is to explore possibilities of using maser emission for investigating massive clusters and associated RSGs. We previously discussed the kinematics of the cluster itself using the velocity information of RSGs with maser emission \citep{nak06}. However, since the number of detection in the present observation is not sufficient enough for discussing cluster kinematics, here we focus on the nature of RSGs. In the following subsections, firstly we compare infrared properties of 4 known RSGs to clarify the difference and similarity between detections and non-detections in the SiO and H$_2$O maser observations. Secondary, we extend our analysis to RSGs in other massive clusters, which previously have been observed in the SiO, H$_2$O and OH maser lines, so that we can more comprehensively consider the characteristics of RSGs with and without maser emission.

\subsection{Comparison between Detections and Non-Detections}

The infrared data used for the present analysis were taken from following data archives: Deep Near Infrared Survey of the Southern Sky \citep[DENIS;][]{deu95}, 2MASS Point Source Catalog \citep[PSC;][]{skr06}, $Spitzer$'s Galactic Legacy Infrared Mid-Plane Survey Extraordinaire \citep[GLIMPSE;][]{wer04}, Midcourse Space Experiment \citep[MSX;][]{ega03}, Wide-field Infrared Survey Explorer (WISE) All Sky Source Catalog \citep{wri10} and AKARI/IRC Point Source Catalog \citep{ish10}. In conjunction with these archival data, we obtained 8--18 photometric data points per each RSG from 0.82~$\mu$m to 22~$\mu$m. The photometric data collected are given in Appendix B.

It has been known that interstellar extinction toward Wd1 is non-negligible in near-infrared wavelengths. Therefore, we adopted an extinction correction for photometric data below 10 $\mu$m in the following way. Firstly, we found $E_{J-K}$ values from \citet{lev05} and \citet{men07} toward each RSG; their $E_{J-K}$ values are based on intrinsic $J$--$K$ colors determined by effective temperatures of theoretical models \citep{lev05,men07}. Secondary, $A_{K}$ values toward each RSG are determined using the formula given in \citet{cro06} [note: some authors recently use a slightly different formula \citep[see, e.g.,][]{nis06}, but the difference of the formulas does not affect the results of our discussion; see, some more details in Section 5.2]. Finally, derived extinction at $K$-band is extrapolated to other wavelengths using the formula given in \citet{ind05}. The averaged extinction toward 4 known RSGs at $K$-band is 1.33. This averaged extinction is consistent with previous studies; for example, the extinction measurements using main-sequence and pre-main-sequence stars in Wd1 shows $A_{K}$ is roughly 1.13 \citep{bra08}, the result of Wolf-Rayet stars shows $A_{K}$ is 1.01 \citep{cro08}, and the results of OB supergiants are 1.20 \citep{cla05} and 1.34 \citep{neg10b}. Both original and interstellar extinction corrected spectral energy distribution (SED) diagrams are given in Figure~3. 

To estimate physical parameters of dust envelopes, SED profiles were fitted by one-dimensional dust radiative transfer models assuming centrally-heated spherical density distributions using the DUSTY code \citep{ive99}. The central star (i.e., central heat source) was assumed to be a point source at the center of the SEDs and their spectral energy distributions were taken to be Planckian. We used the grain type of cold (Sil-Oc) silicates \citep{oss92}. The standard Mathis, Rumpl, Nordsieck \citep[MRN;][]{mat77} power-law was used for the grain size distributions. The dust temperature on the inner shell boundary and the optical depth were varied assuming an inverse square law for the spherical density distribution. The shell was assumed to extend to 10,000 times its inner radius. In order to model the dust shells of RSGs, we surveyed the numerical space of 3 parameters: i.e., the effective temperature of the central heating source ($T_{\rm eff}$), dust temperature on the inner shell boundary ($T_{\rm d}$), and optical depth at 2.2~$\mu$m ($\tau_{2.2}$). The ranges and steps of parameter searches were 1500~K--4000~K and 100~K, 100~K--1800~K and 100K, and 0.001--0.05 and 0.001 for $T_{\rm eff}$, $T_{\rm d}$ and $\tau_{2.2}$, respectively. Additionally, we inspected a couple of models with large optical depths ($\tau_{2.2}>0.05$), but we confirmed, in such a range of large optical depths, that there are no chances to fit the observational SEDs due to the strong absorption of silicate at 9.7~$\mu$m. We adopted the fits for which the sum of squares of the deviations between the observed and modeled fluxes (after scaling) were  minimum. The mass-loss rate calculated by DUSTY posses general scaling properties in terms of the luminosity $L$, gas-to-dust mass ratio $r_{\rm gd}$ and dust grain bulk density $\rho_{\rm s}$, assuming $L=10^{4}$~L$_{\odot}$, $r_{\rm gd}=200$ and $\rho_{\rm s}=3$~g~cm$^{-3}$. The mass-loss rate is scaled in proportion to $L^{3/4} (r_{\rm gd} \rho_{\rm s})^{1/2}$ \citep[This quantity has roughly 30\% inherent uncertainty;][]{ive99}. To obtain the mass-loss rate of each RSG in Wd1,  firstly we calculated the absolute luminosity of each RSG: we integrated flux densities by approximating the area under SED profiles by trapezoids, and the absolute luminosity was derived by assuming the distance of 3.5~kpc. The derived absolute luminosities ($\log L$) are  $5.37$, $6.03$, $5.10$ and $4.83$ for W~237, W~26, W~20 and W~75, respectively. Then, using these absolute luminosities, we scaled the mass-loss rate, assuming $r_{\rm gd}=200$ and $\rho_{\rm s}=3$~g~cm$^{-3}$. Consequently, the mass-loss rates ($\log \dot{M}$) are calculated to be $-5.00$, $-4.65$, $-5.07$ and $-5.69$ for W~237, W~26, W~20 and W~75, respectively. The best-fit input parameters of the DUSTY modeling, obtained mass-loss rates and absolute luminosities used for scaling are summarized in Table 7.

Even though the derived mass-loss rates and luminosities are within a typical range of the mass-loss rate of RSGs \citep{mau11}, W~75, which is a non-detection in both the SiO and H$_2$O maser lines, exhibits the smallest luminosity and mass-loss rate, while two detections (W~237 and W~26) clearly exhibit larger absolute luminosities and mass-loss rates than W~75. According to the results of above analysis, one may expect that RSGs with maser emission are found predominantly in RSGs with relatively large mass-loss rates and absolute luminosities.

\subsection{Consideration using Previous Maser Observations}

As discussed in Section 5.1, the detection rates of maser emission of RSGs seem to be related to infrared properties. However, the number of RSGs in Wd1 is unfortunately not sufficient enough to conclude something. Therefore, in this section, we extend our analysis to RSGs associated with other clusters (i.e., RSGC1, RSGC2, RSGC2 SW, Per OB1, Mc8). Since the galactocentric distance of these 5 clusters are, more or less, similar (except for Per OB1; see Table~6), we can expect that the effects of metallicity gradient in the galactic disk \citep{hen99} are minimized. Unfortunately, at this moment, the distance information is based on radial velocities (i.e., distances are estimated by assuming the flat rotation model of the Galaxy). Therefore, the distances could include relatively large uncertainty [relative error of more than 50\% according to the recent results of trigonometric parallax measurements; Imai, H., in private communication; theoretically, however, the distances to massive clusters may be improved in the future, because there are independent methodologies to measure the distances to clusters \citep[see, e.g.,][]{per98,and07}]. Here, as a preliminary analysis, anyway we discuss based on the kinematic distances estimated from the radial velocities.

In the five massive clusters mentioned above, in total 53 RSGs have been identified until now \citep{hum70,fig06,dav07,deg10}. As well as we did in Section 5.1, we collected infrared photometric data of RSGs from DENIS, 2MASS PSC, $Spitzer$/GLIMPSE, MSX, WISE All Sky Source Catalog and AKARI/IRC PSC. The collected infrared photometric data and SED plots are summarized again in Appendix B. Similarly to the analysis on Wd1, we adopted extinction corrections based on the extinction coefficients given in literature \citep{ind05,lev05,dav07,dav08}. As we mentioned in Section 5.1, some authors recently use a slightly different formula  for the conversion of the extinction values: we used $A_{K}=1.8 \times E_{H-K}$, while some authors recently use $A_{K}=1.5 \times E_{H-K}$ \citep[see, e.g.,][]{nis06}. We numerically evaluated the difference between the results obtained by the two formulas: for the majority of the samples with $A_{K} < 2.5$ (this $A_{K}$ is based on the values given in Table B1), the relative difference in the finally derived luminosities is less than 20\%. For exceptional sources with a large extinction ($A_{K} > 3.0$), the relative difference occasionally goes beyond 40\%, but such objects are quite rare (only two). Therefore, the selection of the conversion formula does not affect the results of our discussion. In Figure~4, we present the distribution of 29 RSGs in 6 clusters (including Wd1) on the two-color diagram of $\log (F_{21}/F_{12})$ and [K--12.13] colors. Although 5 RSGs in Per OB1 (the red open diamonds) exhibit relatively small [K--12.13] values, there seems to be no other significant inhomogeneity in the distribution [since RSGs in Per OB1 have been identified mainly in optical \citep{hum70}, the sample could be somewhat biased]. Then, absolute luminosities and mass-loss rates were estimated using the same methodology as we adopted for Wd1. The obtained absolute luminosity and the best-fit parameters of DUSTY models are summarized in Table 8. Then, the results of previous maser observations of 53 RSGs in 5 clusters are summarized in Tables 9 and 10. 

Figure~5 shows the two-color diagrams of RSGs in 5 clusters and Wd1. The data points in Figure~5 are classified into either detections (red filled circles) or non-detections (blue open circles) in each maser line. In these two-color diagrams, we clearly see that the distributions of maser detections and non-detections are different in colors: i.e., detections are distributed in the upper-right of each panel, while the non-detections are distributed in lower-left of each panel. Since generally $\log (F_{21}/F_{12})$ and [K--12.13] are related to  dust temperature and mid-infrared optical depth of the envelopes, respectively \citep[see, e.g.,][]{sid96}, the concentration of maser detections suggests that maser emission is selectively detected toward RSGs with a low dust temperature and large optical depth. On the other hand, we cannot see clear difference between the SiO, H$_2$O and OH maser lines in the distributions of detected and non-detected sources (even though the number of OH maser observations is not sufficient enough for comparison).

Interestingly, however, one may point out a weak but possible difference between the SiO, H$_2$O and OH maser lines in the distributions of detections and non-detections in the $\log \dot{M}$ versus $\log L$ plot given in Figure~6 (Note: in this figure we do not intend to discuss the correlation between the mass-loss rate and absolute luminosity; it is natural that there is a correlation between $\dot{M}$ and $L$ in Figure~6, because both values are derived from the same SED. The main focus here is only the distributions of detections and non-detections in mass-loss rates and absolute luminosities). The detections in the SiO line distribute in a wide range of $L$, whereas the majority of the H$_2$O maser detections tend to locate in a region with large $L$ (above, roughly $10^{5}$~L$_{\odot}$) except for one data point (St2-08) around $\log L \sim 4.55$. Of course, we need to bear in mind that the distance includes a relatively large uncertainty, and the number of the samples is not large enough. However, as a possibility, one may interpret this difference for astrophysical reasons; for example, difference in excitation conditions of masers, such as, difference in excitation temperatures and critical densities. And if this difference is originated in the intrinsic astrophysical reasons, one may expect that maser lines are used to select a particular group of RSGs lying in a small range of mass-loss rate and luminosity. On the OH maser observation, at the moment we cannot see any tendency, because the number of observations are clearly too small to say something. Since the physical conditions required for pumping OH maser are also different from those for SiO and H$_2$O masers, one may expect that there will be a recognizable difference from SiO and H$_2$O masers if the number of OH observations is increased. 

Using the photometric data collected, we inspected the distribution of detections and non-detections of maser searches on the $Q1$--$Q2$ diagram, which has been suggested by \citet{mes12}. \citet{mes12} proposed a new source selection method for identifying mass-losing evolved stars, which is based on photometric data from 2MASS ($J$, $H$, $K$-band magnitudes) and GLIMPSE (8.0$\mu$m-band magnitude) and on the $Q1$ and $Q2$ parameters. These two parameters are independent of interstellar extinction. The $Q1$ parameter [$Q1=(J-H)-1.8\times (H-K)$] is a measure of the deviation from the reddening vector in the $J-H$ versus $H-K$ plane. Early-type stars have $Q1$ values around 1, K-giants around 0.4 mag, and dusty circumstellar envelopes of evolved stars generate even smaller $Q1$ values. On average, higher mass-loss rates produce smaller $Q1$ values \citep{mes12}. Similarly, the $Q2$ parameter [$Q2=(J-K)-2.69\times(K-[8.0])$; here, [8.0] means the GLIMPSE 8 $\mu$m-band magnitude] measures the deviation from the reddening vector in the plane $J-K$ versus $K-[8.0]$. On average, higher mass-loss rates imply smaller $Q2$ values as well as $Q1$ \citep{mes12}.

In Figure~7, we present the distribution of the present RSG samples on the $Q1$--$Q2$ diagram (red and blue marks); the samples of Mira-like AGB stars \citep{mes02} and Galactic RSGs \citep{ver12} are also plotted together (gray marks) for comparison. Unfortunately, only 13 data points of the present samples can be plotted on the diagram, because for the majority of stars the GLIMPSE 8 $\mu$m data are not available due to source confusion (the present RSG samples are taken from star clusters, in which RSGs roll up in a small region). Nevertheless, as suggested by \citet{ver12}, we see a clear tendency in the diagram: the detections in the SiO $J=1$--0 lines exhibit negative $Q1$ values, while non-detections exhibit relatively larger $Q1$ value. Although we see the detection in the H$_2$O maser line (St2-08) at $(Q1, Q2)=(0.33, 0.61)$, this presumably suggests that the SiO maser line is more sensitive to the $Q$-parameters than the H$_2$O maser line \citep[St2-08 is negative in the SiO maser search,][]{deg10}. In the previous work by \citet{ver12} on Galactic RSG samples, the detections in the SiO $J=2$--1 line are lying below $Q2=-1.9$. However, in the present case we find one detection (Mc8-05) above $Q2=-1.9$. Anyway, a larger statistics is required to confirm the suggested boundary on the $Q1$--$Q2$ diagram.

\section{Summary}
In this paper, we have reported the result of an ATCA observations of the Wd1 region in the SiO $v=1$, $J=1$--$0$ and H$_2$O 6$_{16}$--5$_{23}$ lines. We analyzed infrared SEDs of 57 RSGs in 6 clusters including Wd1 to investigate the correlations between maser and infrared properties. The main results of this research are summarized below:
 
\begin{enumerate}
\item The SiO $v=1$, $J=1$--$0$ and H$_2$O 6$_{16}$--5$_{23}$ lines are detected toward 2 of 4 known RSGs in Wd1, and the large velocity ranges of detected maser lines are consistent with the RSG status. The velocity ranges of the SiO maser line are wider than that of the H$_2$O line; this fact may suggest that the RSG envelopes are deviated from a spherical symmetry.
\item RSGs with maser emission seem to exhibit relatively large $\log (F_{21}/F_{12})$ and [K--12.13] colors compared to non-detections. The mass-loss rates derived from dust radiative transfer modeling suggests that RSGs with maser emission exhibit a relatively large mass-loss rate compared to RSGs with no maser emission. 
\item RSGs with SiO maser emission homogeneously distribute in $L$, whereas those with H$_2$O maser emission tend to distribute in a region with large $L$ values (above, roughly $4 \times 10^{4}$~L$_{\odot}$).
\item The distribution of the detections and non-detection on the $Q1$--$Q2$ diagram is roughly consistent with the previous study by \citet{ver12}, even though one data point of a SiO maser detection is found above $Q2=-1.9$. 
\end{enumerate}

This work is supported by grants awarded to Jun-ichi Nakashima from the Research Grants Council of Hong Kong (project code: HKU 703308P), the Seed Funding Program for Basic Research of the University of Hong Kong (project code: 200802159006) and the Small Project Funding of the University of Hong Kong (project code: 201007176004). The Australia Telescope Compact Array is part of the Australia Telescope National Facility which is funded by the Commonwealth of Australia for operation as a National Facility managed by CSIRO. The authors thank Ryszard Szczerba, Mikako Matsuura and Hiroshi Imai for stimulating discussions and valuable comments.


\appendix

\section{New SiO Maser Detection toward IRAS 16105--4205}

We detected the SiO $v=1$, $J=1$--$0$ line toward IRAS 16105--4205 for the first time. The peak velocity and peak flux are $-$83~km~s$^{-1}$ and 8.8~Jy, respectively. The velocity-integrated flux of the detection is 28.2~Jy~km~s$^{-1}$. The SiO $v=1$, $J=2$--$1$ line has been detected previously by \citet{hai94}. In Figure A1, we present the line profile of the $v=1$, $J=1$--$0$ line superimposed on that of the $v=1$, $J=2$--$1$ line \citep{hai94}. IRAS 16105--4205 was previously classified as a carbon star based on near-infrared photometric colors \citep{fou92}. However, since the object is detected in SiO maser lines, we suspect that it may be an oxygen--rich star rather than a carbon star.  In fact, \citet{dol87} and \citet{che02} respectively detected the OH 1612 MHz maser emission and 9.7~$\mu$m silicate feature toward this object, although a H$_2$O maser search was negative \citep{deg89}.

\section{Photometric Data of RSGs Associated with Massive Clusters}

Photometric data and extinction coefficients used for the analysis are summarized in Tables B1--B5. Spectral energy distribution diagrams and the results of DUSTY modeling are given in Figures~B1--B3.


\clearpage


\begin{figure}
\epsscale{.70}
\plotone{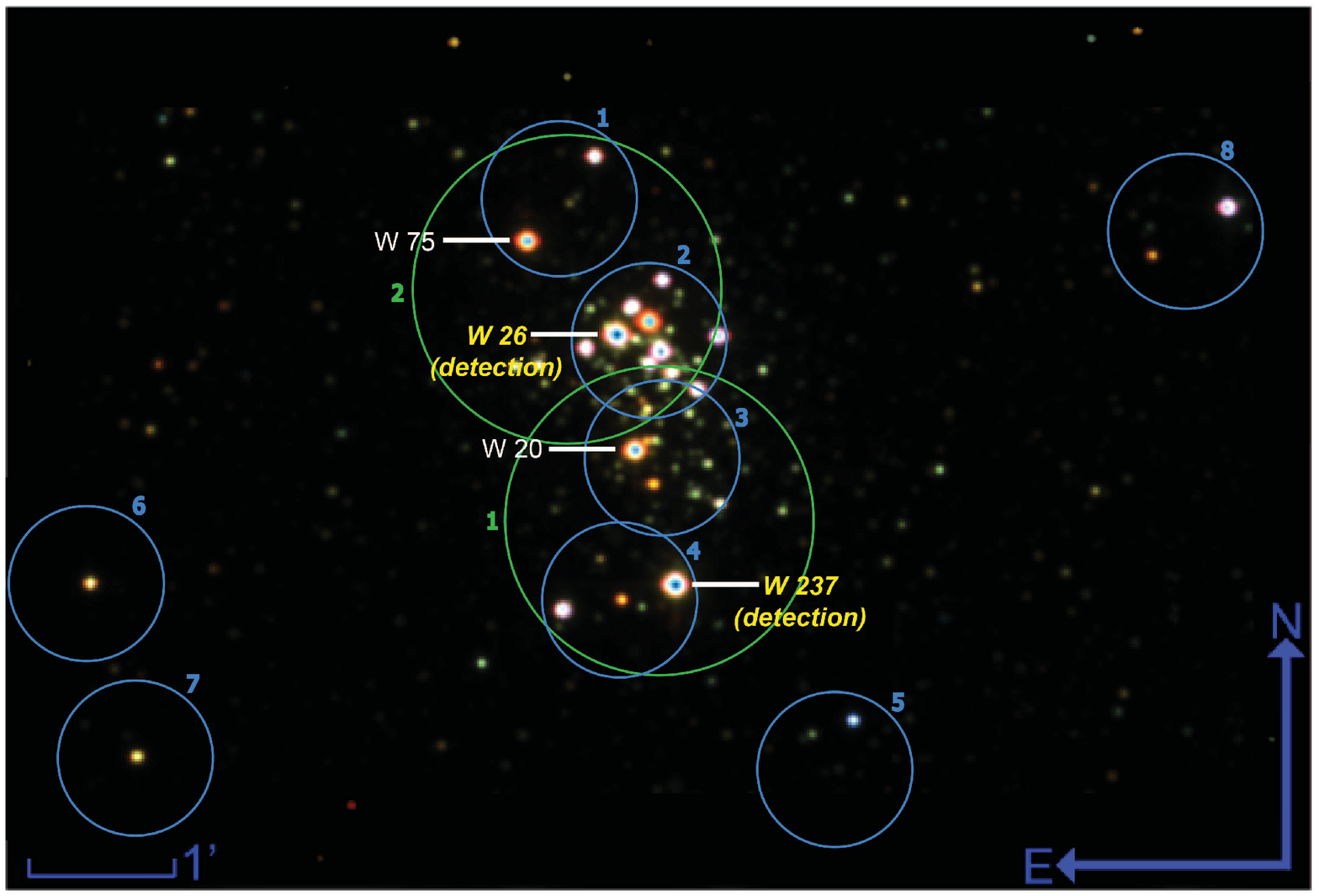}
\caption{Positions of FOVs for the SiO and H$_2$O maser observations at 43~GHz and 22~GHz, respectively, superimposed on the $J$, $H$ and $K$-bands color composite 2MASS image (image size is 8.8$'$ $\times$ 5.9$'$; north is up, east is left). The blue and green circles represent FOVs for the SiO and H$_2$O observations, respectively. The diameters of the blue and green circles (i.e., HPBW of an ATCA 22~m antenna) are 66$''$ and 126$''$, respectively. Each FOV is numbered according to Table~1. Known RSGs with no maser emission are labeled with object names in white color. RSGs detected in the SiO and H$_2$O maser lines are labeled with object names in yellow color. \label{fig1}}
\end{figure}

\clearpage

\begin{figure}
\epsscale{.80}
\plotone{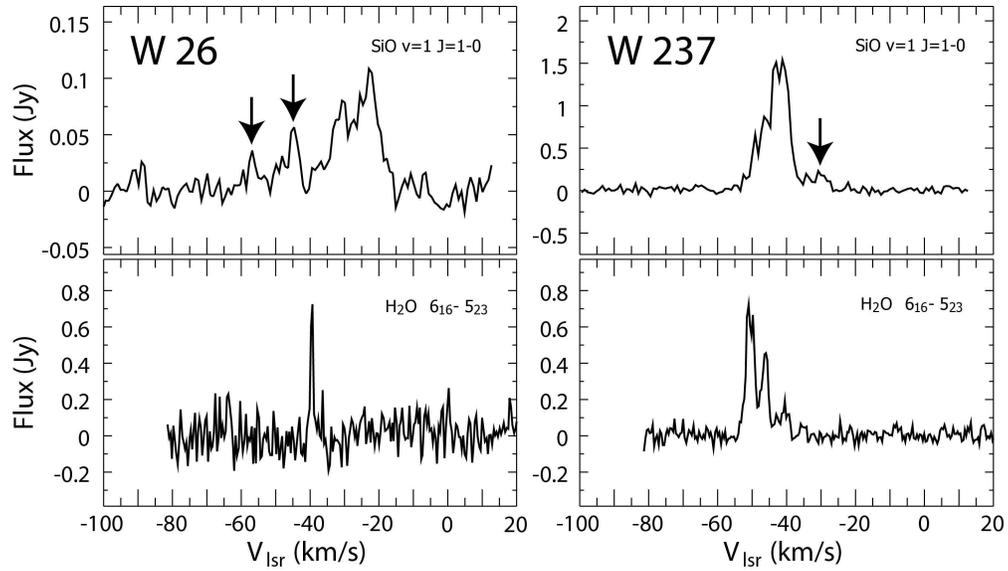}
\caption{Spectra of the SiO $v=1$, $J=1$--$0$ (upper panels) and H$_2$O 6$_{16}$--5$_{23}$ (lower panels) maser lines. The left and right panels show the line profiles of W~26 and W~237, respectively. The arrows indicate weak components ($>5\sigma$) shifted from the primary  peak (see, text). \label{fig2}}
\end{figure}

\clearpage

\begin{figure}
\epsscale{.80}
\plotone{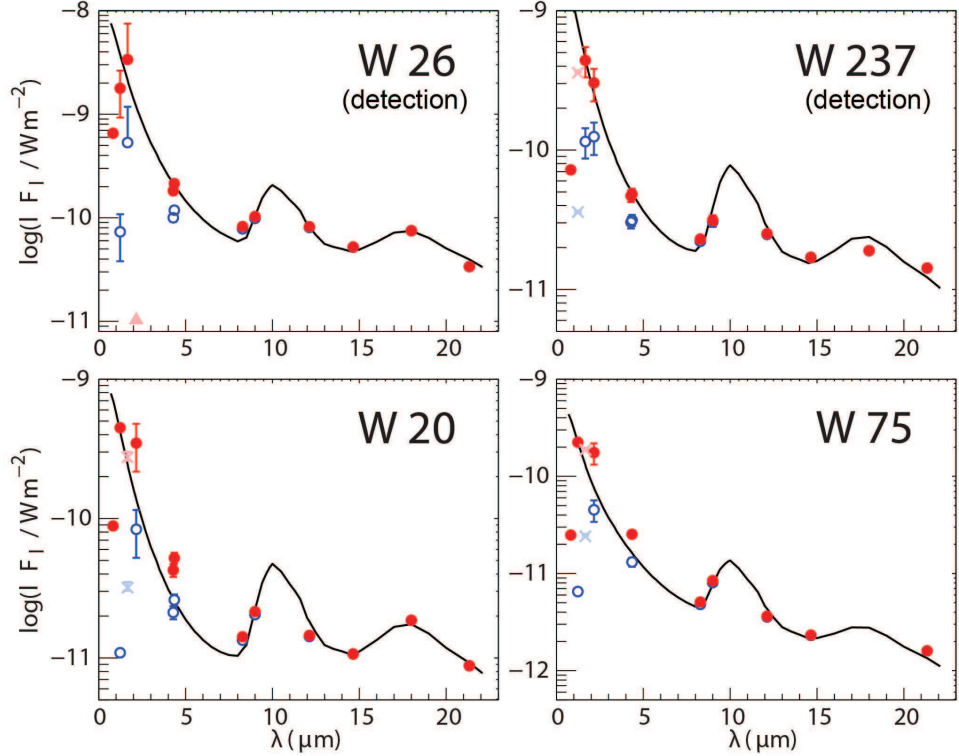}
\caption{Spectral energy distributions of 4 known RSGs in Wd1. The blue open circles are data points of original values without interstellar extinction corrections, and the red filled circles are data points with interstellar extinction corrections. The light-blue and light-red triangles represent the lower limits of fluxes, respectively, before and after adopting interstellar extinction correction, and the light-blue and light-red crosses represent the flux values with a low quality flag, respectively, before and after adopting interstellar extinction correction. The black curves are the results of model fitting with the DUSTY code (see, text for details). \label{fig3}}
\end{figure}

\clearpage

\begin{figure}
\epsscale{.80}
\plotone{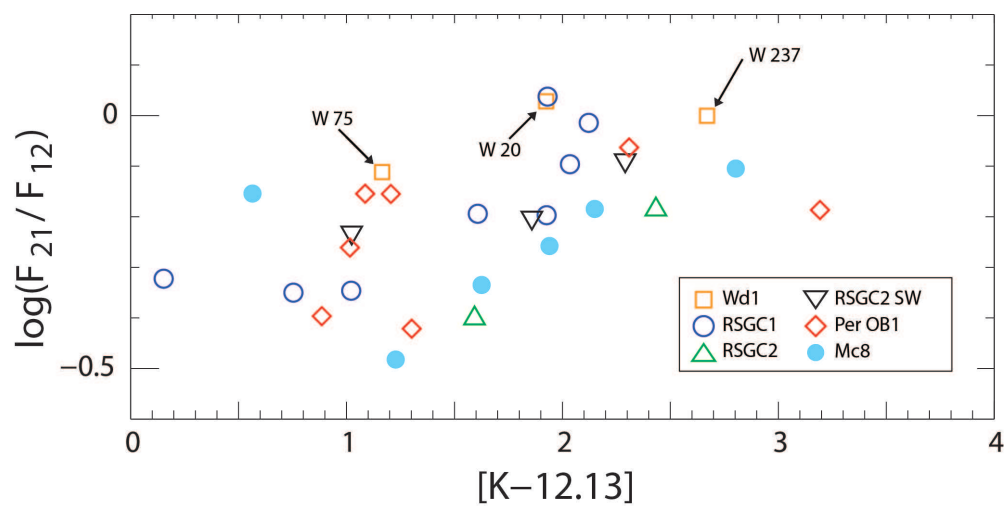}
\caption{Distribution of RSGs in 6 massive clusters on the infrared two-color diagram. The member RSGs of Wd1 are labeled with the arrows (W~26 is not indicated due to a large uncertainty in its K-band flux). In the $\log (F_{21}/F_{12})$ color, $F_{21}$ and $F_{12}$ represent the MSX 21.34~$\mu$m and 12.13~$\mu$m fluxes, respectively. The [K--12.13] color is defined by [K--12.13]=$K+2.5 \log (F_{12}/26.51 {\rm [Jy]})$ \citep{ega99}; here, $K$ and $F_{12}$ are the 2MASS $K$-band magnitude and the MSX 12.13~$\mu$m flux. The interstellar extinction corrections were adopted (see Sections 5.1 and 5.2).  \label{fig4}}
\end{figure}

\clearpage

\begin{figure}
\epsscale{.80}
\plotone{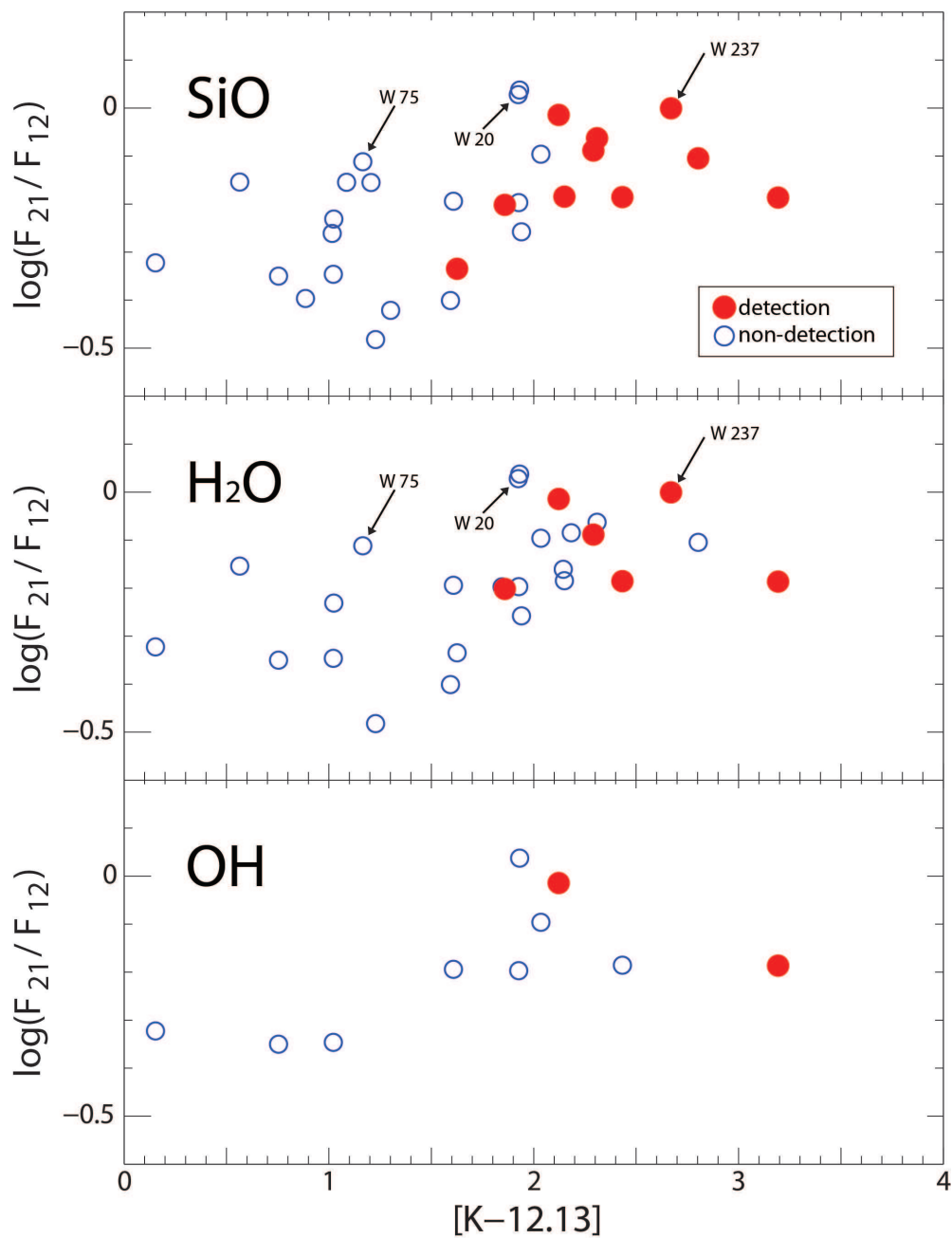}
\caption{Distributions of detections and non-detections of maser observations in the SiO ($J=1$--0, $v=1$ and 2), H$_2$O (22 GHz) and OH maser (1612~MHz) lines on the infrared two-color diagrams. The definitions of colors are the same as Figure~4. The red filled and blue open circles represent detections and non-detections. The member RSGs of Wd1 are indicated by the arrows. W~26 is not indicated in this plot due to large uncertainty in its photometric value. \label{fig5}}
\end{figure}

\clearpage

\begin{figure}
\epsscale{.80}
\plotone{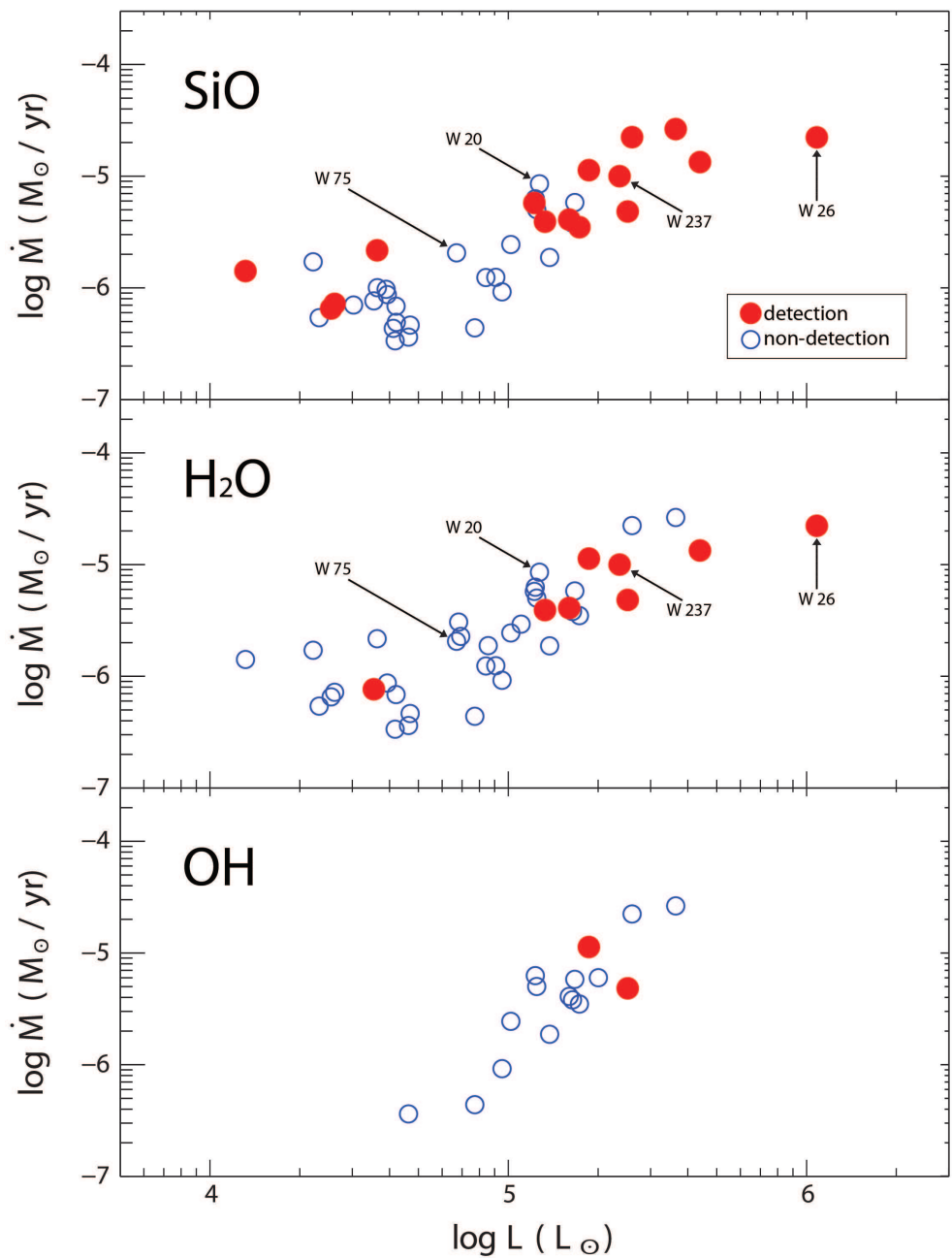}
\caption{Distributions of detections and non-detections of maser observations in the SiO ($J=1$--0, $v=1$ and 2), H$_2$O (22 GHz) and OH maser (1612~MHz) lines on the mass-loss rate versus absolute luminosity diagram. The red filled and blue open circles represent detections and non-detections. The member RSGs of Wd1 are indicated by the arrows. \label{fig6}}
\end{figure}

\clearpage

\begin{figure}
\epsscale{0.6}
\plotone{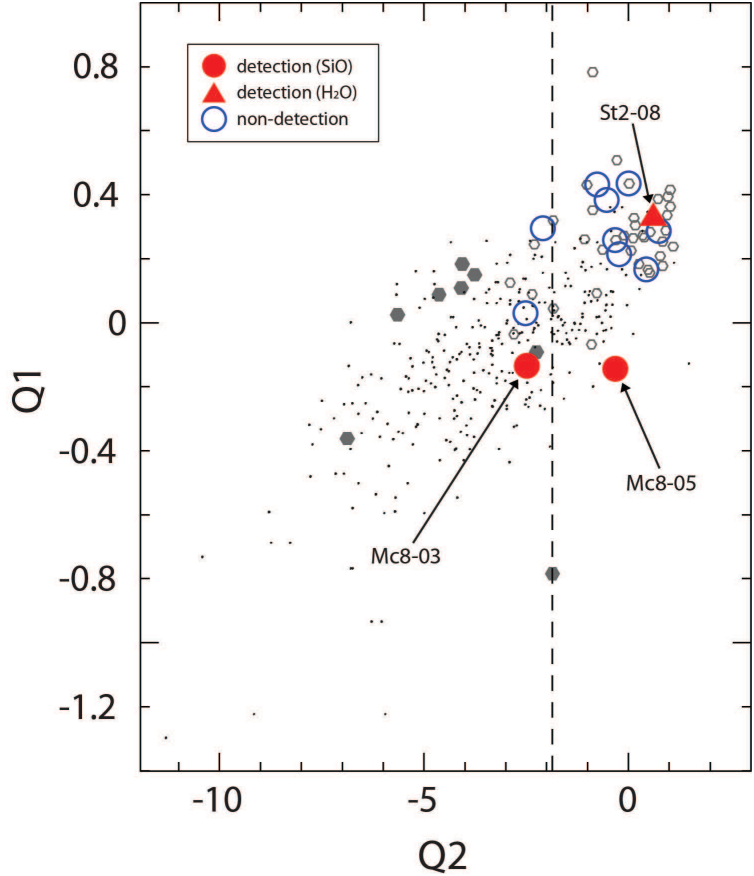}
\caption{Distributions of detections and non-detections of maser observations on the extinction-free $Q1$ and $Q2$ parameters overlaid on a selected region of Figure~3 in \citet{ver12}. The red filled circles represent RSGs detected only in the SiO maser line (and both H$_2$O and OH maser searches are negative). The red filled triangle represents RSG detected only in the H$_2$O maser line (and both SiO and OH maser searches are negative). The open blue circles represent the RSGs that are negative in all three maser lines (SiO, H$_2$O and OH). The gray filled and empty hexagons represent RSGs with and without maser detections, respectively \citep{ver12}. The small dots represent SiO maser detections of Mira-like AGB stars \citep{mes02}. The vertical dotted line represent the Q2 value of $-$1.9 (see, text). RSGs detected in maser lines are indicated by arrows along with object names. \label{fig6}}
\end{figure}

\clearpage

\renewcommand{\thefigure}{A\arabic{figure}}
\setcounter{figure}{0}
\begin{figure}
\epsscale{.60}
\plotone{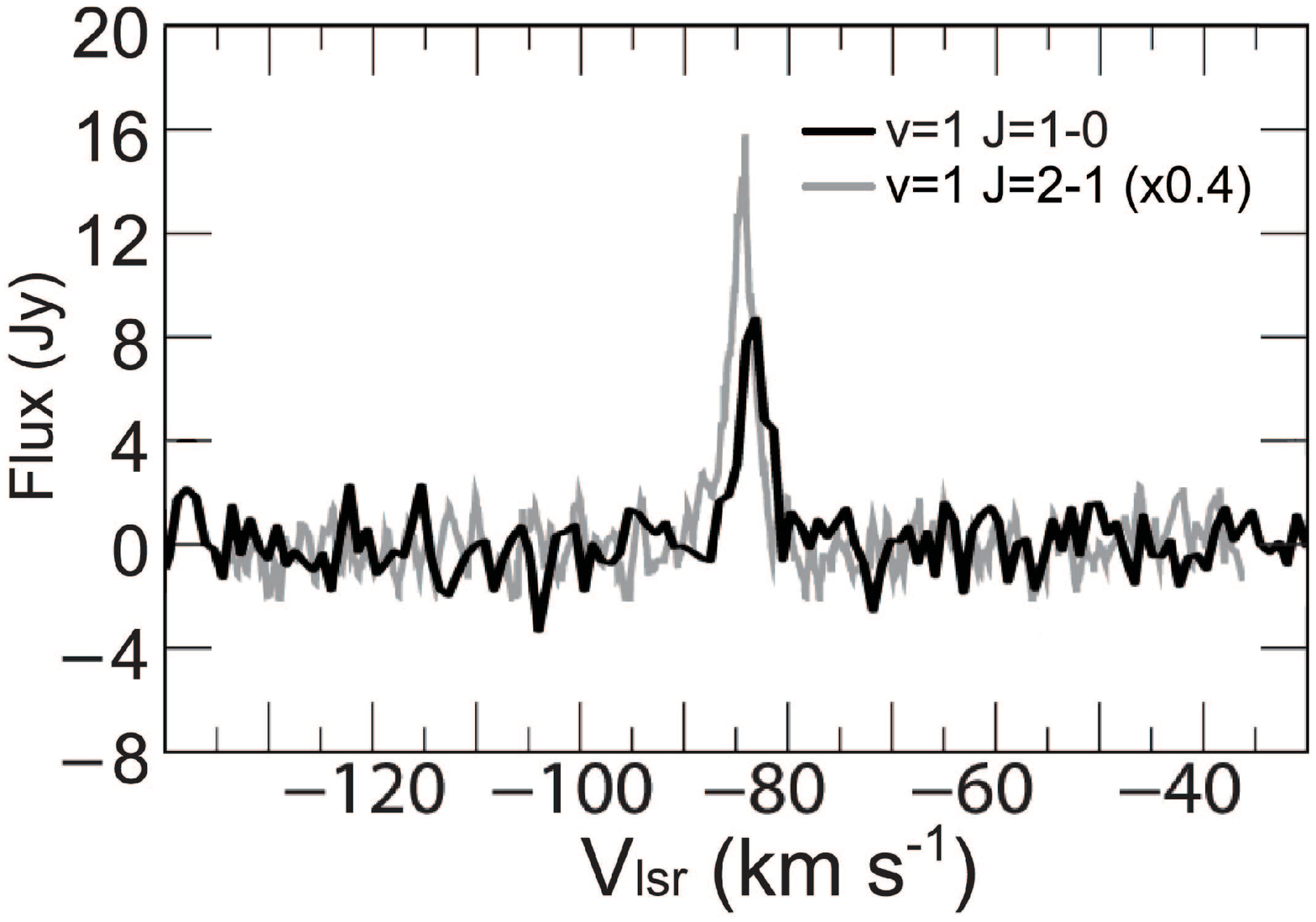}
\caption{Spectrum of the SiO $v=1$, $J=1$--$0$ line of IRAS 16105--4205 (thick line), superimposed on the spectrum of the SiO $v=1$, $J=2$--$1$ line \citep[gray line;][]{hai94}. The $v=1$, $J=1$--$0$ line is detected for the first time. \label{figA1}}
\end{figure}

\clearpage

\renewcommand{\thefigure}{B\arabic{figure}}
\setcounter{figure}{0}
\begin{figure}
\epsscale{.90}
\plotone{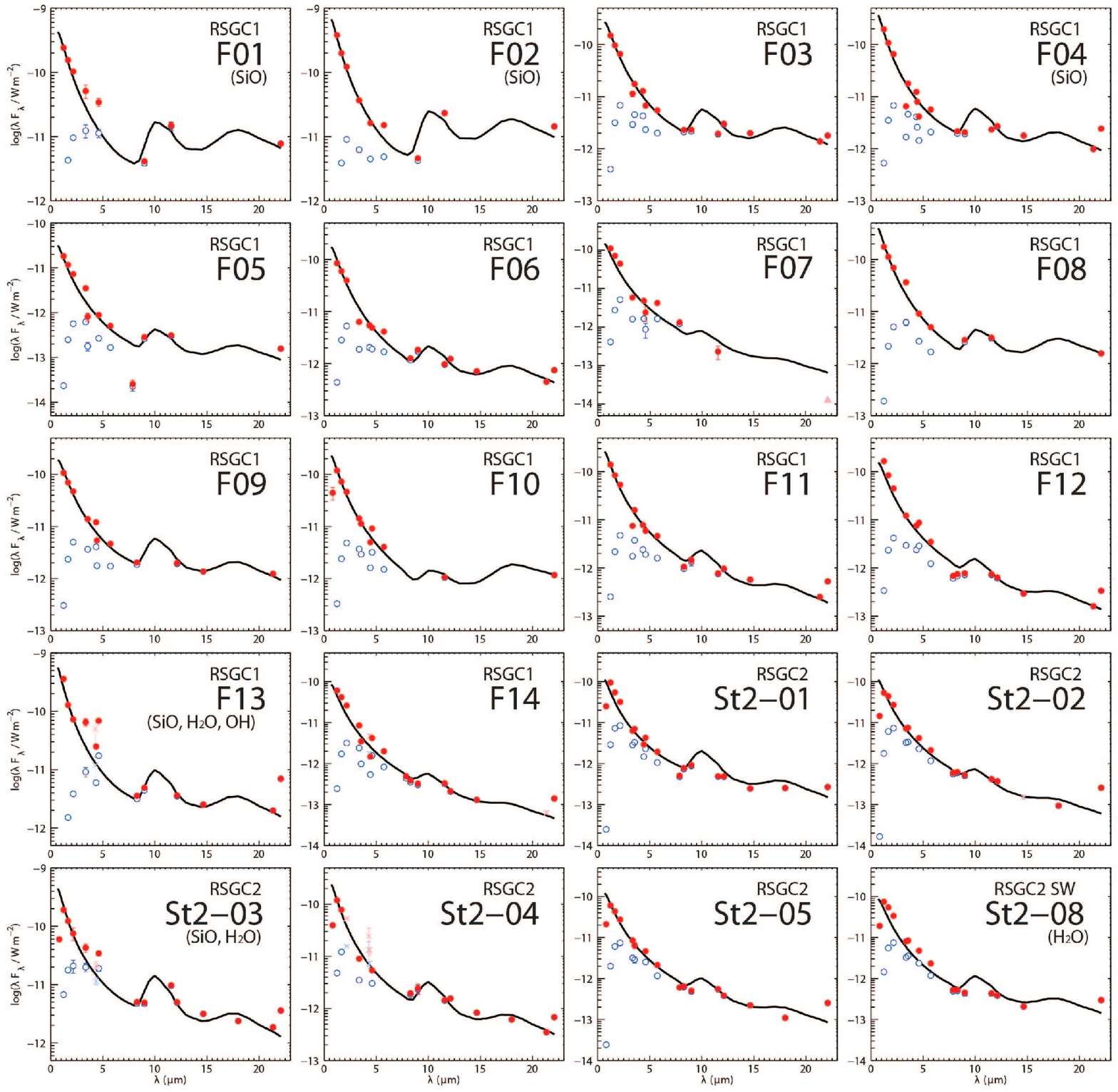}
\caption{Spectral energy distributions of 53 known RSGs in 5 massive clusters. The notations of data points and model curves are the same as Figure~3. The names of parent clusters (small font) and object names (large font) are given in the upper-right corners of each panel. The names of objects with maser detections are indicated in boldface, and detected maser molecules are presented in the parenthesis. \label{figB1}}
\end{figure}

\clearpage

\renewcommand{\thefigure}{B\arabic{figure}}
\begin{figure}
\epsscale{.90}
\plotone{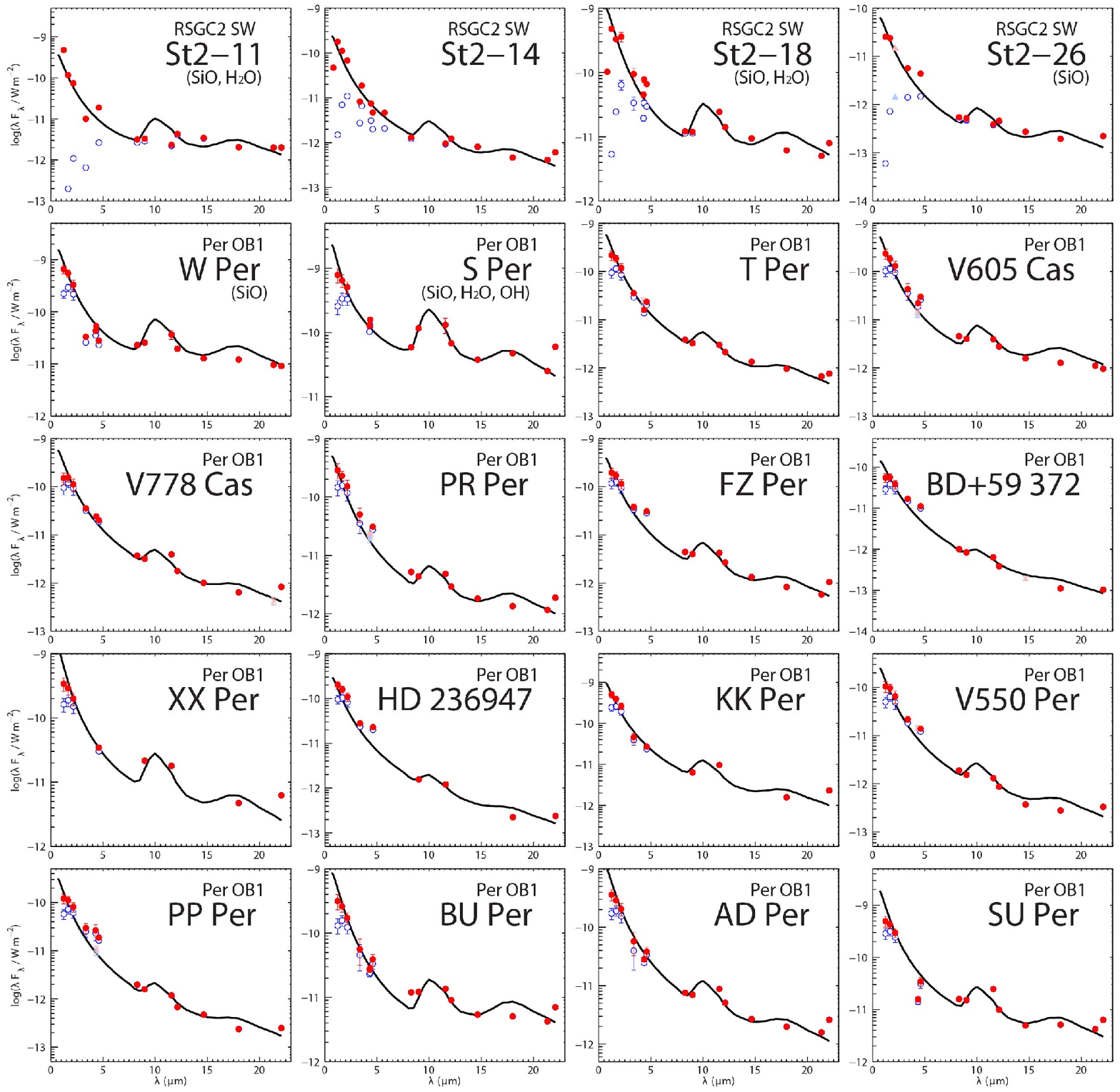}
\caption{Continue of Figure~B1. \label{figB2}}
\end{figure}

\clearpage

\renewcommand{\thefigure}{B\arabic{figure}}
\begin{figure}
\epsscale{.90}
\plotone{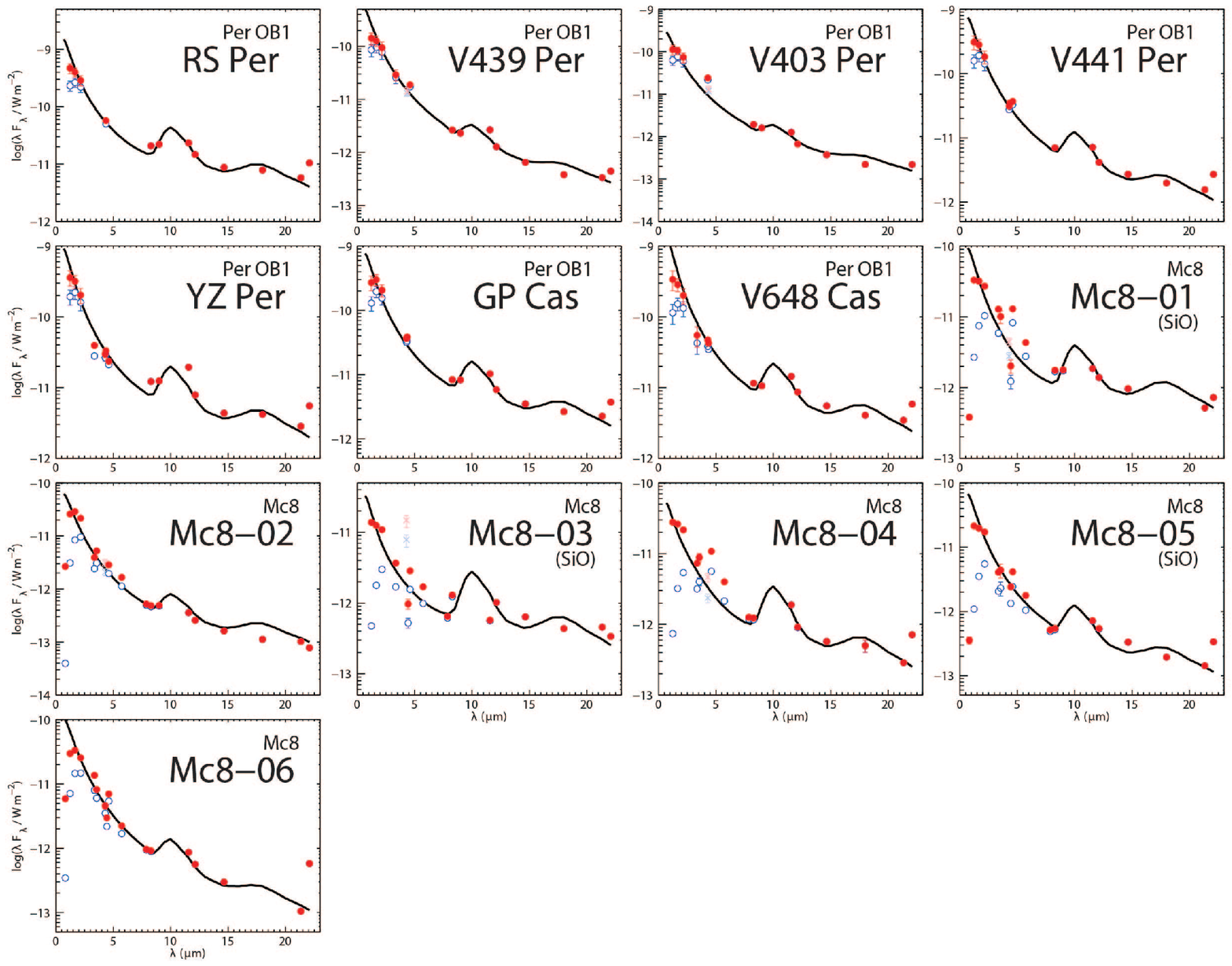}
\caption{Continue of Figure~B2. \label{figB3}}
\end{figure}

\clearpage


\begin{deluxetable}{lccccr}
\tabletypesize{\scriptsize}
\tablecaption{Observational parameters}
\tablewidth{0pt}
\tablehead{\colhead{Position of} & \multicolumn{2}{c}{Center position} & \colhead{Integration time} & \colhead{Beam size\tablenotemark{*}} & \colhead{P.A.\tablenotemark{*}} \\ 
\colhead{observation} & \colhead{R.A. (J2000.0)} & \colhead{Dec. (J2000.0)} & \colhead{(hr)} & \colhead{} & \colhead{}
}
\startdata
\multicolumn{5}{l}{SiO Observation} \\
\tableline 
Position 1 & 16:47:07.65 & $-$45:49:39.3 & 0.65 & 5.2$''$ $\times$ 2.5$''$ & 5.6$^{\circ}$  \\
Position 2 & 16:47:04.20 & $-$45:50:38.3 & 0.99 & 5.1$''$ $\times$ 2.6$''$ & 2.7$^{\circ}$  \\
Position 3 & 16:47:03.62 & $-$45:51:26.3 & 0.99 & 5.0$''$ $\times$ 2.6$''$ & 4.7$^{\circ}$  \\
Position 4 & 16:47:05.34 & $-$45:52:24.3 & 0.50 & 8.1$''$ $\times$ 2.2$''$ & 12.5$^{\circ}$ \\
Position 5 & 16:46:56.82 & $-$45:53:33.2 & 0.50 & 5.9$''$ $\times$ 2.4$''$ & 14.7$^{\circ}$ \\
Position 6 & 16:47:26.12 & $-$45:52:17.3 & 0.50 & 8.5$''$ $\times$ 2.2$''$ & 5.3$^{\circ}$  \\
Position 7 & 16:47:24.31 & $-$45:53:29.3 & 0.50 & 8.3$''$ $\times$ 2.2$''$ & 9.3$^{\circ}$  \\
Position 8 & 16:46:43.06 & $-$45:49:53.0 & 0.50 & 5.8$''$ $\times$ 2.4$''$ & 16.6$^{\circ}$ \\
\tableline \tableline
\multicolumn{5}{l}{H$_2$O Observation} \\
\tableline
Position 1 & 16:47:03.90 & $-$45:51:51.4 & 2.03 & 0.9$''$ $\times$ 0.4$''$ & $-$40.6$^{\circ}$ \\   
Position 2 & 16:47:07.20 & $-$45:50:17.5 & 0.58 & 1.7$''$ $\times$ 0.3$''$ & $-$33.7$^{\circ}$ \\    

\enddata
\tablenotetext{*}{Parameters of synthesized beams.}
\end{deluxetable}

\clearpage

\begin{deluxetable}{lccc}
\tabletypesize{\scriptsize}
\tablecaption{List of known RSGs in Westerlund 1}
\tablewidth{0pt}
\tablecolumns{4}
\tablehead{\colhead{Source} & \colhead{2MASS} & \multicolumn{2}{c}{2MASS position} \\
\colhead{name} & \colhead{name} & \colhead{R.A. (J2000.0)} & \colhead{Dec. (J2000.0)} }
\startdata
W 237 & J16470309$-$4552189 & 16:47:03.09 & $-$45:52:18.9 \\
W 26  & J16470540$-$4550367 & 16:47:05.54 & $-$45:50:36.9 \\
W 20  & J16470468$-$4551238 & 16:47:04.69 & $-$45:51:23.9 \\
W 75  & J16470892$-$4549585 & 16:47:08.93 & $-$45:49:58.6 \\
\enddata
\end{deluxetable}

\clearpage

\begin{deluxetable}{lcccccc}
\tabletypesize{\scriptsize}
\tablecaption{Parameters of detected maser lines}
\tablewidth{0pt}
\tablecolumns{7}
\tablehead{\colhead{Source} & \colhead{RA\tablenotemark{*}} & \colhead{Dec\tablenotemark{*}} & \colhead{$I_{\mathrm{peak}}$} & \colhead{$I_{\mathrm{int}}$} & \colhead{$V_{\mathrm{peak}}$} & \colhead{rms} \\ 
\colhead{} & \colhead{(J2000.0)} & \colhead{(J2000.0)} & \colhead{(Jy)} & \colhead{(Jy km~s$^{-1}$)} & \colhead{(km~s$^{-1}$)} & \colhead{(Jy)} }
\startdata
\sidehead{SiO ($v=1$, $J=1$--$0$)}
\tableline 
W 237 & 16:47:03.07 & $-$45:52:17.8 & 1.54 & 16.82 & $-$41.2 & 0.027 \\
W 26  & 16:47:05.15 & $-$45:50:35.5 & 0.12 & 3.05  & $-$22.9 & 0.018 \\
\tableline \tableline
\sidehead{H$_{2}$O (6$_{16}$--5$_{23}$)}
\tableline
W 237 & 16:47:03.12 & $-$45:52:19.2 & 0.72 & 9.86 & $-$51.0 & 0.028 \\
W 26  & 16:47:05.41 & $-$45:50:36.9 & 0.73 & 3.29 & $-$39.3 & 0.094 \\
\enddata
\tablenotetext{*}{Radio position obtained by fitting two-dimensional Gaussian function.}
\end{deluxetable}

\clearpage

\begin{deluxetable}{lcccccccc}
\tabletypesize{\scriptsize}
\tablecaption{Noise levels toward non-detected RSGs}
\tablewidth{0pt}
\tablecolumns{9}
\tablehead{\colhead{Source} & \colhead{rms (SiO)} & \colhead{rms (H$_{2}$O)}\\ 
\colhead{} & \colhead{(Jy)} & \colhead{(Jy)}}
\startdata
W 20 & 0.005 & 0.028 \\
W 75 & 0.009 & 0.094 \\
\enddata
\end{deluxetable}

\clearpage


\clearpage

\begin{deluxetable}{lccc}
\tabletypesize{\scriptsize}
\tablecaption{Summary of velocity measurements of Westerlund 1 \label{tbl-6}}
\tablewidth{0pt}
\tablecolumns{3}
\tablehead{\colhead{Measurement} & \colhead{Averaged velocity} & \colhead{Reference}\\
\colhead{} & \colhead{(km~s$^{-1}$)} & \colhead{}}
\startdata
SiO maser     & $-$40 & see Section 4\\
H$_2$O maser  & $-$44 & see Section 4\\
CO bandhead   & $-$51 & \citet{men09}\\
CO emission   & $-48\pm14$ & \citet{lun09}\\
HI absorption & $-$55$^{+9}_{-26}$ & \citet{kot07}\\

\enddata

\end{deluxetable}

\clearpage


\clearpage

\begin{deluxetable}{lccrcccc}
\tabletypesize{\scriptsize}
\tablecaption{Parameters of clusters \label{tbl-6}}
\tablewidth{0pt}
\tablecolumns{8}
\tablehead{\colhead{Cluster} & \colhead{Adopted dis.} & \colhead{Galactocentric dis.} & \colhead{Averaged $V_{\mathrm{lsr}}$} & \colhead{Ref.} & \colhead{\begin{tabular}[c]{@{}c@{}}Kin. dis.\\kot07$^{*}$\end{tabular}} & \colhead{\begin{tabular}[c]{@{}c@{}}Kin. dis.\\rei09$^{\dagger}$\end{tabular}} & \colhead{Averaged A$_K$}\\
\colhead{} & \colhead{(kpc)} & \colhead{(kpc)} & \colhead{(km/s)} & \colhead{} & \colhead{(kpc)} & \colhead{(kpc)} & \colhead{(mag)}}
\startdata
Wd1      & 3.5 & 4.5 & $-$39.9~~~~~~ & \tablenotemark{1} & 3.0 & 3.0 & 1.33\\
RSGC1    & 6.6 & 3.3 &   120.9~~~~~~ & \tablenotemark{2} & 6.9 & 6.0 & 2.57\\
RSGC2    & 5.5 & 3.6 &    96.6~~~~~~ & \tablenotemark{3} & 5.2 & 5.1 & 1.40\\
RSGC2 SW & 5.5 & 3.6 &    96.6~~~~~~ & \tablenotemark{3} & 5.2 & 5.2 & 2.52\\
Per OB1  & 2.4 & 9.5 & $-$38.5~~~~~~ & \tablenotemark{4} & 3.3 & 3.0 & 0.31\\
Mc8      & 5.2 & 3.3 &    95.0~~~~~~ & \tablenotemark{3} & 5.2 & 5.2 & 1.11\\
\enddata

\tablenotetext{*}{Calculated based on the Galaxy model of \citet{kot07}}
\tablenotetext{\dagger}{Calculated based on the Galaxy model of \citet{rei09}}

\tablerefs{$^1${see Section 4};~$^2$\citet{dav08};~$^3$\citet{deg10};~$^4$\citet{asa10}}

\end{deluxetable}


\clearpage

\begin{deluxetable}{lccccc}
\tabletypesize{\scriptsize}
\tablecaption{DUSTY model parameters and absolute luminosities}
\tablewidth{0pt}
\tablecolumns{6}
\tablehead{\colhead{Source} & \colhead{$T_{\mathrm{eff}}$} & \colhead{$T_d$} & \colhead{$\tau_{2.2}$} & \colhead{log $\dot{M}$} & \colhead{$\log L$}\\ 
\colhead{name} & \colhead{(K)} & \colhead{(K)} & \colhead{} & \colhead{(M$_{\odot}$~yr$^{-1}$)} & \colhead{(L$_{\odot}$)} }
\startdata
W 237      & 3600 & ~600 & 0.004 & $-$5.00 & 5.37\\
W 26       & 3700 & ~500 & 0.002 & $-$4.65 & 6.03\\
W 20       & 3500 & ~500 & 0.005 & $-$5.07 & 5.10\\
W 75       & 3600 & 1100 & 0.004 & $-$5.69 & 4.83\\
\enddata
\end{deluxetable}

\clearpage

\begin{deluxetable}{lccccc}
\tabletypesize{\scriptsize}
\tablecaption{DUSTY model parameters and absolute luminosities \label{tbl-1}}
\tablewidth{0pt}
\tablecolumns{6}
\tablehead{\colhead{Source} & \colhead{$T_{\mathrm{eff}}$} & \colhead{$T_d$} & \colhead{$\tau_{2.2}$} & \colhead{log $\dot{M}$} & \colhead{$\log L$}\\ 
\colhead{name} & \colhead{(K)} & \colhead{(K)} & \colhead{} & \colhead{(M$_{\odot}$~yr$^{-1}$)} & \colhead{(L$_{\odot}$)} }
\startdata
\sidehead{RSGC1}
\tableline
F01        & 3300 & ~300 & 0.005 & $-$4.65 & 5.41\\
F02        & 3600 & ~300 & 0.004 & $-$4.58 & 5.56\\
F03        & 3200 & ~400 & 0.002 & $-$5.24 & 5.22\\
F04        & 3900 & ~500 & 0.001 & $-$5.46 & 5.24\\
F05        & 3600 & ~400 & 0.001 & $-$5.42 & 5.22\\
F06        & 3300 & ~400 & 0.001 & $-$5.61 & 5.01\\
F07        & 3600 & 1800 & 0.001 & $-$6.36 & 4.89\\
F08        & 3700 & ~300 & 0.001 & $-$5.22 & 5.30\\
F09        & 3100 & ~500 & 0.003 & $-$5.30 & 5.10\\
F10        & 3500 & ~200 & 0.001 & $-$5.20 & 5.09\\
F11        & 3600 & 1200 & 0.002 & $-$5.73 & 5.14\\
F12        & 3200 & 1000 & 0.001 & $-$6.04 & 4.98\\
F13        & 4100 & ~500 & 0.001 & $-$5.32 & 5.40\\
F14        & 3500 & 1500 & 0.001 & $-$6.44 & 4.67\\
\tableline \tableline
\sidehead{RSGC2}
\tableline
St2-01     & 3800 & 1400 & 0.003 & $-$6.06 & 4.59\\
St2-02     & 3500 & 1500 & 0.001 & $-$6.47 & 4.62\\
St2-03     & 3700 & ~900 & 0.003 & $-$5.39 & 5.20\\
St2-04     & 3700 & ~800 & 0.001 & $-$5.91 & 4.93\\
St2-05     & 3400 & 1200 & 0.001 & $-$6.33 & 4.67\\
\tableline \tableline
\sidehead{RSGC2 SW}
\tableline
St2-08     & 3300 & ~600 & 0.001 & $-$6.12 & 4.55\\
St2-11     & 3700 & ~600 & 0.002 & $-$5.41 & 5.12\\
St2-14     & 3500 & ~800 & 0.001 & $-$5.90 & 4.96\\
St2-18     & 3200 & ~500 & 0.003 & $-$4.87 & 5.64\\
St2-26     & 3300 & ~500 & 0.001 & $-$6.14 & 4.42\\
\tableline \tableline
\sidehead{Per OB1}
\tableline
W Per      & 3400 & ~600 & 0.004 & $-$5.24 & 5.09\\
S Per      & 3600 & ~900 & 0.010 & $-$4.95 & 5.27\\
T Per      & 3500 & 1100 & 0.001 & $-$6.31 & 4.62\\
V605 Cas   & 3600 & ~500 & 0.001 & $-$6.00 & 4.56\\
V778 Cas   & 3500 & 1200 & 0.001 & $-$6.36 & 4.61\\
PR Per     & 3300 & ~500 & 0.001 & $-$6.01 & 4.59\\
FZ Per     & 3400 & 1300 & 0.003 & $-$6.15 & 4.48\\
BD +59 372 & 3100 & 1500 & 0.001 & $-$6.93 & 4.08\\
XX Per     & 3700 & ~900 & 0.002 & $-$5.73 & 4.93\\
HD 236947  & 3400 & 1500 & 0.001 & $-$6.69 & 4.35\\
KK Per     & 3500 & 1300 & 0.002 & $-$6.00 & 4.85\\
V550 Per   & 3500 & 1500 & 0.002 & $-$6.51 & 4.27\\
PP Per     & 3700 & 1500 & 0.001 & $-$6.67 & 4.32\\
BU Per     & 3300 & ~400 & 0.002 & $-$5.51 & 4.83\\
AD Per     & 3400 & ~900 & 0.001 & $-$6.01 & 4.90\\
SU Per     & 3700 & ~700 & 0.001 & $-$5.72 & 5.10\\
RS Per     & 3500 & ~900 & 0.003 & $-$5.53 & 5.04\\
V439 Per   & 3600 & 1500 & 0.001 & $-$6.52 & 4.54\\
V403 Per   & 3500 & 1500 & 0.001 & $-$6.70 & 4.31\\
V441 Per   & 3200 & 1000 & 0.002 & $-$5.95 & 4.78\\
YZ Per     & 3400 & ~800 & 0.002 & $-$5.64 & 4.84\\
GP Cas     & 3300 & ~800 & 0.002 & $-$5.84 & 4.78\\
V648 Cas   & 3700 & ~700 & 0.001 & $-$5.79 & 5.01\\
\tableline \tableline
\sidehead{Mc8}
\tableline
Mc8-01     & 3200 & ~600 & 0.004 & $-$5.66 & 4.56\\
Mc8-02     & 3200 & ~600 & 0.001 & $-$6.27 & 4.37\\
Mc8-03     & 3300 & ~900 & 0.010 & $-$5.85 & 4.12\\
Mc8-04     & 3300 & 1400 & 0.015 & $-$5.77 & 4.35\\
Mc8-05     & 3200 & ~900 & 0.002 & $-$6.18 & 4.41\\
Mc8-06     & 3100 & 1500 & 0.003 & $-$6.16 & 4.62\\
\enddata

\end{deluxetable}

\clearpage

\begin{deluxetable}{lcccccc}
\tabletypesize{\scriptsize}
\tablecaption{Summary of previous maser observations of other clusters 1}
\tablewidth{0pt}
\tablecolumns{7}
\tablehead{\colhead{Source} & \colhead{IRAS} & \multicolumn{2}{c}{Coordinate} & \multicolumn{3}{c}{Maser observation} \\ 
\colhead{name} & \colhead{name} & \colhead{RA} & \colhead{Dec} & \colhead{SiO} & \colhead{H$_{2}$O} & \colhead{OH} \\ 
\colhead{} & \colhead{} & \colhead{(J~2000.0)} & \colhead{(J~2000.0)} & \colhead{} & \colhead{} & \colhead{} }
\startdata
\sidehead{RSGC1}
\tableline
F01        & -- & 18:37:56.30 & $-$06:52:32.2 & y\tablenotemark{1} & n\tablenotemark{1} & n\tablenotemark{2} \\
F02        & -- & 18:37:55.28 & $-$06:52:48.3 & y\tablenotemark{1} & n\tablenotemark{1} & n\tablenotemark{2} \\
F03        & -- & 18:37:59.73 & $-$06:53:49.2 & n\tablenotemark{1} & n\tablenotemark{1} & n\tablenotemark{2} \\
F04        & -- & 18:37:50.88 & $-$06:53:38.1 & y\tablenotemark{1} & n\tablenotemark{1} & n\tablenotemark{2} \\
F05        & -- & 18:37:55.52 & $-$06:52:11.6 & n\tablenotemark{1} & n\tablenotemark{1} & n\tablenotemark{2} \\
F06        & -- & 18:37:57.45 & $-$06:53:25.3 & n\tablenotemark{1} & n\tablenotemark{1} & n\tablenotemark{2} \\
F07        & -- & 18:37:54.31 & $-$06:52:34.5 & n\tablenotemark{1} & n\tablenotemark{1} & n\tablenotemark{2} \\
F08        & -- & 18:37:55.19 & $-$06:52:10.7 & --                 & --                 & n\tablenotemark{2} \\
F09        & -- & 18:37:57.76 & $-$06:52:22.1 & n\tablenotemark{1} & n\tablenotemark{1} & n\tablenotemark{2} \\
F10        & -- & 18:37:59.53 & $-$06:53:31.9 & n\tablenotemark{1} & n\tablenotemark{1} & n\tablenotemark{2} \\
F11        & -- & 18:37:51.72 & $-$06:51:49.7 & n\tablenotemark{1} & n\tablenotemark{1} & n\tablenotemark{2} \\
F12        & -- & 18:38:03.30 & $-$06:52:45.1 & n\tablenotemark{1} & n\tablenotemark{1} & n\tablenotemark{2} \\
F13        & -- & 18:37:58.91 & $-$06:52:32.1 & y\tablenotemark{1} & y\tablenotemark{1} & y\tablenotemark{2} \\
F14        & -- & 18:37:47.65 & $-$06:53:02.1 & n\tablenotemark{1} & n\tablenotemark{1} & n\tablenotemark{2} \\
\tableline \tableline
\sidehead{RSGC2}
\tableline
St2-01     & --               & 18:39:19.88 & $-$06:01:48.0 & n\tablenotemark{3} & n\tablenotemark{3} & -- \\
St2-02     & --               & 18:39:18.25 & $-$06:02:14.2 & n\tablenotemark{3} & n\tablenotemark{3} & -- \\
St2-03     & IRAS 18366--0603 & 18:39:19.59 & $-$06:00:41.6 & y\tablenotemark{3} & y\tablenotemark{3} & n\tablenotemark{4} \\
St2-04     & --               & 18:39:24.60 & $-$06:02:13.9 & n\tablenotemark{3} & n\tablenotemark{3} & -- \\
St2-05     & --               & 18:39:14.70 & $-$06:01:36.5 & n\tablenotemark{3} & n\tablenotemark{3} & -- \\
\tableline \tableline
\sidehead{RSGC2 SW}
\tableline
St2-08     & -- & 18:39:07.75 & $-$06:03:20.2 & n\tablenotemark{3} & y\tablenotemark{3} & -- \\
St2-11     & -- & 18:39:05.59 & $-$06:04:26.6 & y\tablenotemark{3} & y\tablenotemark{3} & -- \\
St2-14     & -- & 18:39:08.04 & $-$06:05:24.2 & n\tablenotemark{3} & n\tablenotemark{3} & -- \\
St2-18     & -- & 18:39:02.38 & $-$06:05:10.6 & y\tablenotemark{3} & y\tablenotemark{3} & -- \\
St2-26     & -- & 18:38:56.98 & $-$06:06:45.7 & y\tablenotemark{3} & n\tablenotemark{3} & -- \\
\tableline \tableline
\sidehead{Per OB1}
\tableline
W Per      & IRAS 02469+5646 & 02:50:37.91 & 56:59:00.1 & y\tablenotemark{5} & n\tablenotemark{6} & -- \\
S Per      & IRAS 02192+5821 & 02:22:51.73 & 58:35:11.2 & y\tablenotemark{7} & y\tablenotemark{8} & y\tablenotemark{9} \\
T Per      & IRAS 02157+5843 & 02:19:21.87 & 58:57:40.3 & n\tablenotemark{10} & -- & -- \\
V605 Cas   & IRAS 02167+5926 & 02:20:22.48 & 59:40:16.8 & n\tablenotemark{10} & -- & -- \\
V778 Cas   & IRAS 01550+5901 & 01:58:28.90 & 59:16:08.7 & n\tablenotemark{10} & -- & -- \\
PR Per     & IRAS 02181+5738 & 02:21:42.41 & 57:51:46.0 & n\tablenotemark{10} & -- & -- \\
FZ Per     & IRAS 02174+5655 & 02:20:59.65 & 57:09:30.0 & n\tablenotemark{10} & -- & -- \\
BD +59 372 & IRAS 01561+6000 & 01:59:39.67 & 60:15:01.9 & -- & -- & -- \\
XX Per     & IRAS 01597+5459 & 02:03:09.36 & 55:13:56.6 & -- & n\tablenotemark{6} & -- \\
HD 236947  & IRAS 02036+5832 & 02:07:12.06 & 58:47:15.9 & -- & -- & -- \\
KK Per     & IRAS 02068+5619 & 02:10:15.79 & 56:33:32.7 & -- & -- & -- \\
V550 Per   & IRAS 02116+5754 & 02:15:13.30 & 58:08:32.3 & -- & -- & -- \\
PP Per     & IRAS 02135+5817 & 02:17:08.23 & 58:31:47.0 & -- & -- & -- \\
BU Per     & IRAS 02153+5711 & 02:18:53.30 & 57:25:16.8 & -- & n\tablenotemark{6} & -- \\
AD Per     & IRAS 02169+5645 & 02:20:29.00 & 56:59:35.2 & -- & -- & -- \\
SU Per     & IRAS 02185+5622 & 02:22:06.89 & 56:36:14.9 & -- & -- & -- \\
RS Per     & IRAS 02188+5652 & 02:22:24.30 & 57:06:34.4 & -- & n\tablenotemark{6} & -- \\
V439 Per   & IRAS 02196+5658 & 02:23:11.03 & 57:11:58.3 & -- & -- & -- \\
V403 Per   & --              & 02:23:24.11 & 57:12:43.1 & -- & -- & -- \\
V441 Per   & IRAS 02217+5712 & 02:25:21.86 & 57:26:14.1 & -- & -- & -- \\
YZ Per     & IRAS 02347+5649 & 02:38:25.42 & 57:02:46.2 & -- & n\tablenotemark{6} & -- \\
GP Cas     & IRAS 02360+5922 & 02:39:50.44 & 59:35:51.3 & -- & -- & -- \\
V648 Cas   & IRAS 02473+5738 & 02:51:03.95 & 57:51:19.9 & -- & -- & -- \\
\tableline \tableline
\sidehead{Mc8}
\tableline
Mc8-01     & IRAS 18258--1058 & 18:28:35.32 & $-$10:56:36.6 & y\tablenotemark{3} & n\tablenotemark{3} & -- \\
Mc8-02     & --               & 18:28:39.13 & $-$10:55:52.7 & n\tablenotemark{3} & n\tablenotemark{3} & -- \\
Mc8-03     & --               & 18:28:52.08 & $-$10:57:57.7 & y\tablenotemark{3} & n\tablenotemark{3} & -- \\
Mc8-04     & --               & 18:28:54.04 & $-$10:56:45.5 & n\tablenotemark{3} & n\tablenotemark{3} & -- \\
Mc8-05     & --               & 18:28:58.46 & $-$10:56:07.0 & y\tablenotemark{3} & n\tablenotemark{3} & -- \\
Mc8-06     & --               & 18:29:03.05 & $-$10:53:15.4 & n\tablenotemark{3} & n\tablenotemark{3} & -- \\
\enddata

\tablerefs{$^1$\citet{nak06};$^2$\citet{dav08};$^3$\citet{deg10};$^4$\citet{wil72};$^5$\citet{jia96};$^6$\citet{tak01};$^7$\citet{cho96};$^8$\citet{pal93};$^9$\citet{szy10};$^{10}$\citet{jia99}}

\end{deluxetable}


\clearpage


\clearpage

\begin{deluxetable}{lccc}
\tabletypesize{\scriptsize}
\tablecaption{Summary of previous maser observations of other clusters 2 \label{tbl-9}}
\tablewidth{0pt}
\tablecolumns{4}
\tablehead{\colhead{Cluster} & \colhead{Line} & \colhead{Detection/No. of RSGs observed} & \colhead{Telescope \& Ref.} \\ }
\startdata
RSGC1    & SiO $v=1$, $J=1$--$0$               & 4$/$13 & NRO45m\tablenotemark{1} \\
         & SiO $v=2$, $J=1$--$0$               & 1$/$13 & NRO45m\tablenotemark{1} \\
         & H$_2$O (6$_{16}$--5$_{23}$)         & 1$/$14 & NRO45m\tablenotemark{1} \\
         & OH $^2\Pi_{3/2}$~$J=3/2$~$F=1$--$2$ & 1$/$14 & VLA\tablenotemark{2} \\
\tableline
RSGC2    & SiO $v=1$, $J=1$--$0$               & 1$/$5 & NRO45m\tablenotemark{3} \\
         & SiO $v=2$, $J=1$--$0$               & 0$/$5 & NRO45m\tablenotemark{3} \\
         & H$_2$O (6$_{16}$--5$_{23}$)         & 1$/$5 & NRO45m\tablenotemark{3} \\
         & OH $^2\Pi_{3/2}$~$J=3/2$~$F=1$--$2$ & 0$/$1 & NRAO43m\tablenotemark{4} \\
\tableline
RSGC2 SW & SiO $v=1$, $J=1$--$0$               & 3$/$5 & NRO45m\tablenotemark{3} \\
         & SiO $v=2$, $J=1$--$0$               & 2$/$5 & NRO45m\tablenotemark{3} \\
         & H$_2$O (6$_{16}$--5$_{23}$)         & 3$/$5 & NRO45m\tablenotemark{3} \\
\tableline
Per OB1  & SiO $v=1$, $J=1$--$0$               & 2$/$7 & NRO45m\tablenotemark{5,6,7} \\
         & SiO $v=2$, $J=1$--$0$               & 1$/$7 & NRO45m\tablenotemark{5,6,7} \\
         & SiO $v=3$, $J=1$--$0$               & 0$/$1 & NRO45m\tablenotemark{6} \\
         & H$_2$O (6$_{16}$--5$_{23}$)         & 1$/$7 & Medicina32m\tablenotemark{8},KSRC34m\tablenotemark{9} \\
         & OH $^2\Pi_{3/2}$~$J=3/2$~$F=1$--$2$ & 1$/$1 & NRT\tablenotemark{10} \\
         & OH $^2\Pi_{3/2}$~$J=3/2$~$F=2$--$2$ & 1$/$1 & NRT\tablenotemark{10} \\
\tableline
Mc8      & SiO $v=1$, $J=1$--$0$               & 3$/$6 & NRO45m\tablenotemark{3} \\
         & SiO $v=2$, $J=1$--$0$               & 3$/$6 & NRO45m\tablenotemark{3} \\
         & H$_2$O (6$_{16}$--5$_{23}$)         & 0$/$6 & NRO45m\tablenotemark{3} \\
\enddata

\tablerefs{$^1$\citet{nak06};$^2$\citet{dav08};$^3$\citet{deg10};$^4$\citet{wil72};$^5$\citet{jia96};$^6$\citet{cho96};$^7$\citet{jia99};$^8$\citet{pal93};$^9$\citet{tak01};$^{10}$\citet{szy10};}

\hspace{-180pt}NRO45m -- 45m radio telescope, Nobeyama Radio Observatory;

\hspace{-152pt}VLA -- The Very Large Array, National Radio Astronomy Observatory;

\hspace{-160pt}Medicina32m -- Medicina 32m Antenna, Istituto di Radioastronomia;

\hspace{-155pt}KSRC34m -- Kashima 34m Antenna, Kashima Space Research Center;

\hspace{-217pt}NRT -- Nan\c{c}ay radio telescope, Observatoire de Paris;

\hspace{-140pt}NRAO43m -- 43-m radiotelescope, National Radio Astronomy Observatory

\end{deluxetable}


\clearpage


\clearpage
\renewcommand{\thetable}{B\arabic{table}}
\setcounter{table}{0}
\begin{deluxetable}{lc@{\extracolsep{0.1in}}rr@{\extracolsep{0.1in}}rrrrrr}
\tabletypesize{\scriptsize}
\tablecaption{Infrared Photometric Data 1\label{tbl-10}}
\tablewidth{0pt}
\tablecolumns{10}
\tablehead{\colhead{Source} & \colhead{$A_K$} & \multicolumn{2}{c}{DENIS$^{a}$} & \multicolumn{6}{c}{2MASS$^{b}$} \\
\cline{2-2}\cline{3-4}\cline{5-10} 
\colhead{name} & \colhead{2.17~$\mu$m} & \colhead{$I$} & \colhead{$\sigma_{I}$} & \colhead{$J$} & \colhead{$\sigma_{J}$} & \colhead{$H$} & \colhead{$\sigma_{H}$} & 
\colhead{$K$} & \colhead{$\sigma_{K}$} \\
\colhead{}       & \colhead{(mag)} & \colhead{(mag)} & \colhead{(mag)}    & \colhead{(mag)} & \colhead{(mag)}    & \colhead{(mag)} & \colhead{(mag)}    & 
\colhead{(mag)} & \colhead{(mag)}}
\startdata
\sidehead{Wd1}
\tableline 
W 237      & 0.96 & 10.25 & 0.03~~~ & 5.08:  & 0.02~ & 3.01~ & 0.27~ & 2.18~         & 0.28~ \\
W 26       & 1.33 &  9.74 & 0.04~~~ & 4.31~  & 0.52~ & 1.35~ & 1.32~ & 6.21$\dagger$ & --  ~ \\
W 20       & 1.55 & 13.05 & 0.03~~~ & 6.38~  & 0.02~ & 4.38: & 0.08~ & 2.61~         & 0.41~ \\
W 75       & 1.47 & 14.03 & 0.03~~~ & 6.93~  & 0.02~ & 4.67: & 0.02~ & 3.28~         & 0.27~ \\
\tableline \tableline
\sidehead{RSGC1}
\tableline
F01        & 2.58 & --    & --  ~~~ &  9.75~ & 0.03~ & 6.59~ & 0.02~ & 4.96~ & 0.02~ \\
F02        & 2.83 & --    & --  ~~~ &  9.90~ & 0.03~ & 6.70~ & 0.02~ & 5.03~ & 0.02~ \\
F03        & 2.46 & --    & --  ~~~ &  9.95~ & 0.03~ & 6.92~ & 0.02~ & 5.33~ & 0.02~ \\
F04        & 2.46 & --    & --  ~~~ &  9.66~ & 0.03~ & 6.80~ & 0.02~ & 5.34~ & 0.02~ \\
F05        & 2.77 & --    & --  ~~~ & 10.55~ & 0.03~ & 7.18~ & 0.03~ & 5.54~ & 0.03~ \\
F06        & 2.19 & --    & --  ~~~ &  9.87~ & 0.03~ & 7.04~ & 0.03~ & 5.61~ & 0.03~ \\
F07        & 2.33 & --    & --  ~~~ &  9.94~ & 0.03~ & 7.07~ & 0.03~ & 5.63~ & 0.02~ \\
F08        & 2.84 & --    & --  ~~~ & 10.77~ & 0.03~ & 7.33~ & 0.02~ & 5.65~ & 0.02~ \\
F09        & 2.44 & --    & --  ~~~ & 10.26~ & 0.03~ & 7.24~ & 0.02~ & 5.67~ & 0.02~ \\
F10        & 2.45 & 18.47 & 0.29~~~ & 10.18~ & 0.03~ & 7.22~ & 0.03~ & 5.71~ & 0.02~ \\
F11        & 2.63 & --    & --  ~~~ & 10.47~ & 0.03~ & 7.33~ & 0.03~ & 5.72~ & 0.02~ \\
F12        & 2.57 & --    & --  ~~~ & 10.14~ & 0.03~ & 7.24~ & 0.02~ & 5.86~ & 0.03~ \\
F13        & 3.19 & --    & --  ~~~ & 10.91~ & 0.03~ & 7.72~ & 0.02~ & 5.96~ & 0.02~ \\
F14        & 2.29 & --    & --  ~~~ & 10.50~ & 0.03~ & 7.58~ & 0.02~ & 6.17~ & 0.02~ \\
\tableline \tableline
\sidehead{RSGC2}
\tableline
St2-01     & 1.45 & 13.92 & 0.03~~~ &  7.82~ & 0.02~ & 6.02~ & 0.03~ & 5.11~ & 0.02~ \\
St2-02     & 1.42 & 14.35 & 0.03~~~ &  8.35~ & 0.03~ & 6.21~ & 0.03~ & 5.26~ & 0.02~ \\
St2-03     & 1.39 & 12.66 & 0.03~~~ &  6.90~ & 0.02~ & 5.05~ & 0.02~ & 4.12~ & 0.27~ \\
St2-04     & 1.34 & 12.85 & 0.03~~~ &  7.27~ & 0.02~ & 5.46~ & 0.03~ & 4.50: & 0.04~ \\
St2-05     & 1.42 & 13.94 & 0.03~~~ &  8.22~ & 0.02~ & 6.21~ & 0.02~ & 5.24~ & 0.02~ \\
\tableline \tableline
\sidehead{RSGC2 SW}
\tableline
St2-08     & 1.64 & 15.19 & 0.05~~~ &  8.57~ & 0.02~ & 6.31~ & 0.02~ & 5.23~         & 0.02~ \\
St2-11     & 4.58 & --    & --  ~~~ & 14.23~ & 0.04~ & 9.92~ & 0.02~ & 7.32~         & 0.02~ \\
St2-14     & 1.99 & 16.03 & 0.07~~~ &  8.53~ & 0.02~ & 6.05~ & 0.03~ & 4.82~         & 0.02~ \\
St2-18     & 1.87 & 14.56 & 0.04~~~ &  7.15~ & 0.03~ & 4.70~ & 0.02~ & 2.90~         & 0.19~ \\
St2-26     & 2.52 & --    & --  ~~~ & 12.03~ & 0.04~ & 8.52~ & 0.03~ & 6.95$\dagger$ & --  ~ \\
\tableline \tableline
\sidehead{Per OB1}
\tableline
W Per      & 0.45 & --    & --  ~~~ &  3.10~ & 0.20~ & 2.00~ & 0.17~ & 1.57~ & 0.25~ \\
S Per      & 0.47 & --    & --  ~~~ &  2.95~ & 0.28~ & 1.85~ & 0.24~ & 1.12~ & 0.22~ \\
T Per      & 0.35 & --    & --  ~~~ &  4.03~ & 0.25~ & 3.02~ & 0.17~ & 2.58~ & 0.24~ \\
V605 Cas   & 0.35 & --    & --  ~~~ &  3.95~ & 0.26~ & 3.04~ & 0.20~ & 2.48~ & 0.27~ \\
V778 Cas   & 0.19 & --    & --  ~~~ &  4.03~ & 0.30~ & 2.97~ & 0.21~ & 2.49~ & 0.32~ \\
PR Per     & 0.28 & --    & --  ~~~ &  3.56~ & 0.31~ & 2.68~ & 0.19~ & 2.25~ & 0.30~ \\
FZ Per     & 0.22 & --    & --  ~~~ &  3.80~ & 0.25~ & 2.91~ & 0.20~ & 2.48~ & 0.24~ \\
BD+59 372  & 0.28 & --    & --  ~~~ &  5.33~ & 0.25~ & 4.20~ & 0.23~ & 3.74~ & 0.26~ \\
XX Per     & 0.31 & --    & --  ~~~ &  3.44~ & 0.26~ & 2.48~ & 0.23~ & 1.97~ & 0.24~ \\
HD 236947  & 0.31 & --    & --  ~~~ &  4.02~ & 0.23~ & 3.13~ & 0.19~ & 2.62~ & 0.24~ \\
KK Per     & 0.31 & --    & --  ~~~ &  3.00~ & 0.20~ & 2.14~ & 0.17~ & 1.68~ & 0.21~ \\
V550 Per   & 0.31 & --    & --  ~~~ &  4.72~ & 0.27~ & 3.66~ & 0.20~ & 3.19~ & 0.32~ \\
PP Per     & 0.31 & --    & --  ~~~ &  4.57~ & 0.25~ & 3.52~ & 0.22~ & 2.95~ & 0.24~ \\
BU Per     & 0.36 & --    & --  ~~~ &  3.68~ & 0.29~ & 2.68~ & 0.20~ & 2.19~ & 0.23~ \\
AD Per     & 0.31 & --    & --  ~~~ &  3.38~ & 0.24~ & 2.48~ & 0.21~ & 1.94~ & 0.26~ \\
SU Per     & 0.23 & --    & --  ~~~ &  2.82~ & 0.25~ & 1.93~ & 0.18~ & 1.46~ & 0.22~ \\
RS Per     & 0.29 & --    & --  ~~~ &  3.05~ & 0.22~ & 2.11~ & 0.19~ & 1.56~ & 0.21~ \\
V439 Per   & 0.21 & --    & --  ~~~ &  4.13~ & 0.29~ & 3.22~ & 0.23~ & 2.69~ & 0.30~ \\
V403 Per   & 0.24 & --    & --  ~~~ &  4.48~ & 0.27~ & 3.46~ & 0.22~ & 2.99~ & 0.29~ \\
V441 Per   & 0.28 & --    & --  ~~~ &  3.47~ & 0.26~ & 2.47~ & 0.22~ & 2.04~ & 0.24~ \\
YZ Per     & 0.26 & --    & --  ~~~ &  3.26~ & 0.26~ & 2.30~ & 0.23~ & 1.91~ & 0.26~ \\
GP Cas     & 0.31 & --    & --  ~~~ &  3.69~ & 0.27~ & 2.44~ & 0.21~ & 1.94~ & 0.24~ \\
V648 Cas   & 0.45 & --    & --  ~~~ &  3.83~ & 0.34~ & 2.71~ & 0.22~ & 2.11~ & 0.27~ \\
\tableline \tableline
\sidehead{Mc8}
\tableline
Mc8-01     & 1.05 & 16.37 & 0.09~~~ &  7.90~ & 0.03~ & 5.98~ & 0.03~ & 4.88~ & 0.02~ \\
Mc8-02     & 0.88 & 13.41 & 0.05~~~ &  7.74~ & 0.02~ & 5.86~ & 0.04~ & 4.97~ & 0.02~ \\
Mc8-03     & 1.40 & --    & --  ~~~ &  9.78~ & 0.02~ & 7.54~ & 0.04~ & 6.23~ & 0.02~ \\
Mc8-04     & 1.51 & --    & --  ~~~ &  9.30~ & 0.02~ & 6.90~ & 0.04~ & 5.59~ & 0.02~ \\
Mc8-05     & 1.24 & 17.44 & 0.14~~~ &  8.88~ & 0.04~ & 6.80~ & 0.04~ & 5.57~ & 0.02~ \\
Mc8-06     & 0.59 & 11.06 & 0.02~~~ &  6.84~ & 0.03~ & 5.26~ & 0.03~ & 4.49~ & 0.02~ \\
\enddata

\tablenotetext{a}{For saturating sources, the measured fluxes are calibrated by PSF-fitting.} 
\tablenotetext{b}{For saturating sources, the measured fluxes are calibrated by 1-D radial profiles.} 
\tablenotetext{:}{The colon represents uncertain detection.}
\tablenotetext{\dagger}{The dagger represents 95$\%$ confidence upper limit.}
\end{deluxetable}

\clearpage


\clearpage
\renewcommand{\thetable}{B\arabic{table}}
\begin{deluxetable}{lrr@{\extracolsep{0.1in}}rrrrrr}
\tabletypesize{\scriptsize}
\tablecaption{Infrared Photometric Data 2\label{tbl-10}}
\tablewidth{0pt}
\tablecolumns{9}
\tablehead{\colhead{Source} & \multicolumn{8}{c}{WISE$^{a}$} \\
\cline{2-9} 
\colhead{name}   & \colhead{3.4~$\mu$m} & \colhead{$\sigma_{3.4\micron}$} & \colhead{4.6~$\mu$m} & \colhead{$\sigma_{4.6\micron}$} & 
\colhead{12~$\mu$m} & \colhead{$\sigma_{12\micron}$} & \colhead{22~$\mu$m} & \colhead{$\sigma_{22\micron}$} \\
\colhead{}       & \colhead{(mag)} & \colhead{(mag)}   & \colhead{(mag)} & \colhead{(mag)}   & \colhead{(mag)} & \colhead{(mag)}   & \colhead{(mag)} & \colhead{(mag)}   }
\startdata
\sidehead{RSGC1}
\tableline
F01        & 3.37 & 0.25 & 2.49 & 0.16 & $-$0.71 & 0.16 & $-$2.10          & 0.02~ \\
F02        & 3.88 & 0.09 & --   & --   & $-$1.22 & 0.07 & $-$2.77          & 0.01~ \\
F03        & 4.93 & 0.13 & 4.20 & 0.11 &    1.50 & 0.07 & $-$0.50          & 0.07~ \\
F04        & 5.54 & 0.08 & 4.72 & 0.05 &    1.27 & 0.06 & $-$0.83          & 0.02~ \\
F05        & 3.89 & 0.12 & 4.05 & 0.02 &    0.99 & 0.07 & $-$0.36          & 0.02~ \\
F06        & 5.41 & 0.06 & 4.42 & 0.04 &    2.24 & 0.02 & $-$0.17          & 0.02~ \\
F07        & 5.57 & 0.11 & 5.26 & 0.45 &    3.78 & 0.42 &    4.89$\dagger$ & --  ~ \\
F08        & 3.89 & 0.12 & 4.05 & 0.02 &    0.99 & 0.07 & $-$0.36          & 0.02~ \\
F09        & --   & --   & --   & --   &    --   & --   &    --            & --  ~ \\
F10        & 4.68 & 0.09 & 3.86 & 0.06 &    2.14 & 0.03 & $-$0.04          & 0.02~ \\
F11        & 5.49 & 0.06 & 4.42 & 0.05 &    2.50 & 0.03 &    0.82          & 0.04~ \\
F12        & 4.92 & 0.06 & 3.97 & 0.04 &    2.55 & 0.02 &    1.31          & 0.02~ \\
F13        & 3.47 & 0.16 & 2.02 & 0.08 &    --   & --   & $-$1.98          & 0.01~ \\
F14        & 5.15 & 0.05 & 4.63 & 0.04 &    3.43 & 0.02 &    2.26          & 0.03~ \\
\tableline \tableline
\sidehead{RSGC2}
\tableline
St2-01     & 4.96 & 0.07 & 4.22 & 0.05 &    2.98 & 0.02 &    1.56 & 0.03~ \\
St2-02     & 4.81 & 0.07 & 4.22 & 0.04 &    3.14 & 0.03 &    1.61 & 0.03~ \\
St2-03     & 2.85 & 0.18 & 1.93 & 0.11 & $-$0.26 & 0.03 & $-$1.26 & 0.02~ \\
St2-04     & 4.51 & 0.09 & 3.91 & 0.05 &    1.81 & 0.02 &    0.55 & 0.03~ \\
St2-05     & 4.62 & 0.08 & 4.11 & 0.05 &    2.85 & 0.02 &    1.61 & 0.03~ \\
\tableline \tableline
\sidehead{RSGC2 SW}
\tableline
St2-08     & 4.82 & 0.06 & 4.19 & 0.05 &    3.12 & 0.02 &    1.45 & 0.09~ \\
St2-11     & 6.33 & 0.05 & 4.07 & 0.06 &    1.33 & 0.07 & $-$0.63 & 0.03~ \\
St2-14     & 5.00 & 0.05 & 4.37 & 0.04 &    2.27 & 0.02 &    0.66 & 0.04~ \\
St2-18     & 2.28 & 0.25 & 1.44 & 0.01 & $-$1.27 & 0.06 & $-$2.12 & 0.02~ \\
St2-26     & 5.72 & 0.05 & 4.69 & 0.04 &    3.25 & 0.02 &    1.76 & 0.04~ \\
\tableline \tableline
\sidehead{Per OB1}
\tableline
W Per      & 2.56 & 0.03 & 1.70 & 0.01 & $-$1.73 & 0.21 & $-$2.89 & 0.001 \\
S Per      & --   & --   & --   & --   & $-$3.11 & 0.29 & $-$4.32 & 0.001 \\
T Per      & 2.43 & 0.12 & 1.85 & 0.07 &    1.01 & 0.01 & $-$0.19 & 0.01~ \\
V605 Cas   & 2.21 & 0.32 & 1.58 & 0.09 &    0.71 & 0.02 & $-$0.44 & 0.02~ \\
V778 Cas   & 2.33 & 0.06 & 1.94 & 0.01 &    0.70 & 0.01 &    0.32 & 0.01~ \\
PR Per     & 2.01 & 0.29 & 1.52 & 0.04 &    0.49 & 0.01 & $-$0.56 & 0.02~ \\
FZ Per     & 2.27 & 0.16 & 1.49 & 0.12 &    0.62 & 0.01 &    0.07 & 0.01~ \\
BD+59 372  & 3.19 & 0.13 & 2.62 & 0.05 &    2.69 & 0.01 &    2.60 & 0.02~ \\
XX Per     & --   & --   & 1.41 & 0.05 & $-$0.94 & 0.04 & $-$1.86 & 0.01~ \\
HD 236947  & 2.66 & 0.12 & 1.86 & 0.09 &    1.99 & 0.01 &    1.68 & 0.02~ \\
KK Per     & 2.12 & 0.27 & 1.68 & 0.07 & $-$0.27 & 0.01 & $-$0.79 & 0.01~ \\
V550 Per   & 2.94 & 0.14 & 2.41 & 0.09 &    1.90 & 0.01 &    1.33 & 0.01~ \\
PP Per     & 2.60 & 0.26 & 2.08 & 0.12 &    2.01 & 0.02 &    1.64 & 0.03~ \\
BU Per     & 1.94 & 0.48 & 1.30 & 0.20 & $-$0.64 & 0.02 & $-$1.99 & 0.004 \\
AD Per     & 1.88 & 0.47 & 1.30 & 0.19 & $-$0.17 & 0.02 & $-$0.91 & 0.01~ \\
SU Per     & --   & --   & 1.38 & 0.21 & $-$1.29 & 0.02 & $-$1.88 & 0.01~ \\
RS Per     & --   & --   & --   & --   & $-$1.23 & 0.03 & $-$2.43 & 0.01~ \\
V439 Per   & 2.56 & 0.26 & 2.03 & 0.07 &    1.13 & 0.01 &    1.01 & 0.01~ \\
V403 Per   & --   & --   & --   & --   &    1.94 & 0.02 &    1.78 & 0.02~ \\
V441 Per   & --   & --   & 1.33 & 0.05 &    0.06 & 0.01 & $-$0.95 & 0.01~ \\
YZ Per     & 2.26 & 0.01 & 1.81 & 0.02 & $-$1.03 & 0.01 & $-$1.72 & 0.01~ \\
GP Cas     & --   & --   & --   & --   & $-$0.34 & 0.02 & $-$1.30 & 0.01~ \\
V648 Cas   & 2.02 & 0.34 & --   & --   & $-$0.70 & 0.01 & $-$1.79 & 0.004 \\
\tableline \tableline
\sidehead{Mc8}
\tableline
Mc8-01     & 3.95 & 0.08 & 2.83 & 0.06 & 1.52    & 0.02 & 0.48    & 0.02~ \\
Mc8-02     & 5.14 & 0.06 & 4.39 & 0.04 & 3.31    & 0.02 & 2.28    & 0.03~ \\
Mc8-03     & 5.53 & 0.06 & 4.64 & 0.03 & 2.82    & 0.02 & 1.31    & 0.02~ \\
Mc8-04     & 4.84 & 0.07 & 3.24 & 0.05 & 1.50    & 0.02 & 0.49    & 0.03~ \\
Mc8-05     & 5.31 & 0.06 & 4.16 & 0.04 & 2.57    & 0.02 & 1.31    & 0.03~ \\
Mc8-06     & 3.62 & 0.09 & 3.29 & 0.05 & 2.35    & 0.02 & 0.73    & 0.02~ \\
\enddata

\tablenotetext{a}{For saturating sources, the measured fluxes are calibrated by non-saturated wings of their profiles.}
\tablenotetext{\dagger}{The dagger represents 95$\%$ confidence upper limit.}
\end{deluxetable}

\clearpage


\clearpage
\renewcommand{\thetable}{B\arabic{table}}
\begin{deluxetable}{lrr@{\extracolsep{0.1in}}rrrrrr}
\tabletypesize{\scriptsize}
\tablecaption{Infrared Photometric Data 3\label{tbl-10}}
\tablewidth{0pt}
\tablecolumns{9}
\tablehead{\colhead{Source} & \multicolumn{8}{c}{GLIMPSE$^{a}$} \\
\cline{2-9}
           \colhead{name}   & \colhead{3.6~$\mu$m} & \colhead{$\sigma_{3.6\micron}$} & \colhead{4.5~$\mu$m} & \colhead{$\sigma_{4.5\micron}$} & 
\colhead{5.8~$\mu$m} & \colhead{$\sigma_{5.8\micron}$} & \colhead{8.0~$\mu$m} & \colhead{$\sigma_{8.0\micron}$} \\
           \colhead{}       & \colhead{(mag)} & \colhead{(mag)}         & \colhead{(mag)} & \colhead{(mag)}         & \colhead{(mag)} & \colhead{(mag)}         & \colhead{(mag)} & \colhead{(mag)}         }
\startdata
\sidehead{RSGC1}
\tableline
F01        & --   & --   & --   & --   & --   & --   & --   & --   \\
F02        & --   & --   & 3.54 & 0.07 & 2.69 & 0.05 & --   & --   \\
F03        & 4.23 & 0.11 & --   & --   & 3.64 & 0.04 & --   & --   \\
F04        & 4.21 & 0.04 & 4.13 & 0.05 & 3.60 & 0.04 & --   & --   \\
F05        & 5.24 & 0.22 & --   & --   & 3.85 & 0.03 & 5.05 & 0.24 \\
F06        & --   & --   & --   & --   & 3.83 & 0.03 & --   & --   \\
F07        & --   & --   & 4.62 & 0.07 & 3.86 & 0.03 & 3.25 & 0.04 \\
F08        & --   & --   & --   & --   & 3.89 & 0.03 & --   & --   \\
F09        & 4.47 & 0.05 & 4.55 & 0.07 & 3.80 & 0.03 & --   & --   \\
F10        & 4.70 & 0.03 & 4.64 & 0.04 & 3.96 & 0.03 & --   & --   \\
F11        & 4.43 & 0.10 & --   & --   & 3.89 & 0.03 & --   & --   \\
F12        & --   & --   & --   & --   & 4.17 & 0.03 & 3.99 & 0.04 \\
F13        & --   & --   & --   & --   & --   & --   & --   & --   \\
F14        & 5.89 & 0.12 & 5.88 & 0.12 & 4.65 & 0.03 & 4.36 & 0.03 \\
\tableline \tableline
\sidehead{RSGC2}
\tableline
St2-01     & 4.65 & 0.08 & 4.74 & 0.07 & 4.34 & 0.03 & 4.26 & 0.03 \\
St2-02     & 4.56 & 0.05 & --   & --   & 4.23 & 0.03 & 4.11 & 0.03 \\
St2-03     & --   & --   & --   & --   & --   & --   & --   & --   \\
St2-04     & --   & --   & --   & --   & --   & --   & --   & --   \\
St2-05     & 4.72 & 0.19 & --   & --   & 4.23 & 0.03 & 4.02 & 0.03 \\
\tableline \tableline
\sidehead{RSGC2 SW}
\tableline
St2-08     & 4.56 & 0.06 & --   & --   & 4.22 & 0.02 & 4.22 & 0.03 \\
St2-11     & --   & --   & --   & --   & --   & --   & --   & --   \\
St2-14     & 3.89 & 0.08 & 3.98 & 0.05 & 3.61 & 0.03 & --   & --   \\
St2-18     & --   & --   & --   & --   & --   & --   & --   & --   \\
St2-26     & --   & --   & --   & --   & --   & --   & --   & --   \\
\tableline \tableline
\sidehead{Mc8}
\tableline
Mc8-01     & 4.02 & 0.21 & 4.95 & 0.24 & 3.30 & 0.04 & --   & --   \\
Mc8-02     & 4.65 & 0.06 & --   & --   & 4.27 & 0.03 & 4.21 & 0.02 \\
Mc8-03     & --   & --   & 5.91 & 0.17 & 4.46 & 0.03 & 3.98 & 0.02 \\
Mc8-04     & 4.43 & 0.14 & --   & --   & 3.58 & 0.03 & 3.27 & 0.04 \\
Mc8-05     & 5.03 & 0.26 & 4.85 & 0.07 & 4.35 & 0.03 & 4.22 & 0.03 \\
Mc8-06     & 3.99 & 0.06 & 4.31 & 0.05 & 3.83 & 0.03 & 3.54 & 0.03 \\
\enddata

\tablenotetext{a}{For saturating sources, the measured fluxes are calibrated by non-saturated wings of their profiles. The aperture corrections of point sources were made using the effective correction factors given by Reach et al. (2005).}

\end{deluxetable}

\clearpage


\clearpage

\begin{deluxetable}{lrrrrrrrrrrrr}
\tabletypesize{\scriptsize} \rotate
\tablecaption{Infrared Photometric Data 4\label{tbl-11}}
\tablecolumns{13}
\tablewidth{0pt}
\tablehead{\colhead{Source} & \multicolumn{12}{c}{$MSX$}  \\
\cline{2-13}  
\colhead{name}   & \colhead{4.29~$\mu$m} & \colhead{$\sigma_{4.29\micron}$} & \colhead{4.35~$\mu$m} & \colhead{$\sigma_{4.35\micron}$} & 
\colhead{8.28~$\mu$m} & \colhead{$\sigma_{8.28\micron}$} & \colhead{12.13~$\mu$m} & \colhead{$\sigma_{12.13\micron}$} & 
\colhead{14.65~$\mu$m} & \colhead{$\sigma_{14.65\micron}$} & \colhead{21.34~$\mu$m} & \colhead{$\sigma_{21.34\micron}$} \\
\colhead{}       & \colhead{(Jy)} & \colhead{(Jy)}     & \colhead{(Jy)} & \colhead{(Jy)}     & \colhead{(Jy)} & \colhead{(Jy)}   & \colhead{(Jy)} & \colhead{(Jy)}   & \colhead{(Jy)} & \colhead{(Jy)}   & \colhead{(Jy)} & \colhead{(Jy)}   }
\startdata
\sidehead{Wd1}
\tableline 
W 237      &  43.40~ &  4.21 &  45.50~ &  4.19 &  61.17 & 2.51 & 100.60 &  5.03 &  82.95~ &  5.06 & 101.39~ &  6.08 \\
W 26       & 143.27~ & 12.18 & 171.67~ & 15.11 & 216.37 & 8.87 & 326.91 & 16.35 & 254.14~ & 15.50 & 240.82~ & 14.45 \\
W 20       &  30.33~ &  3.25 &  37.82~ &  3.56 &  37.06 & 1.52 &  57.67 &  2.88 &  52.18~ &  3.18 &  62.56~ &  3.75 \\
W 75       & --    ~ & --    &  19.05~ &  2.04 &  13.27 & 0.54 &  14.43 &  0.72 &  11.25~ &  0.69 &  11.35~ &  0.68 \\
\tableline \tableline
\sidehead{RSGC1}
\tableline
F01        & --    ~ & --    & --    ~ & --    & --     & --   & --     & --    & --    ~ & --    & --    ~ & --   \\
F02        & --    ~ & --    & --    ~ & --    & --     & --   & --     & --    & --    ~ & --    & --    ~ & --   \\
F03        & --    ~ & --    &   6.19~ &  0.59 &   5.77 & 0.24 &  11.91 &  0.60 &   9.68~ &  0.59 &   9.82~ & 0.59 \\
F04        & --    ~ & --    &   5.96~ &  0.57 &   5.42 & 0.22 &  10.71 &  0.54 &   8.75~ &  0.53 &   6.99~ & 0.42 \\
F05        & --    ~ & --    & --    ~ & --    & --     & --   & --     & --    & --    ~ & --    & --    ~ & --   \\ 
F06        & --    ~ & --    &   2.98~ &  0.34 &   3.21 & 0.13 &   4.98 &  0.25 &   3.45~ &  0.21 &   3.19~ & 0.19 \\
F07        & --    ~ & --    & --    ~ & --    & --     & --   & --     & --    & --    ~ & --    & --    ~ & --   \\
F08        & --    ~ & --    & --    ~ & --    & --     & --   & --     & --    & --    ~ & --    & --    ~ & --   \\
F09        & --    ~ & --    &   5.90~ &  0.57 &   5.14 & 0.21 &   7.82 &  0.39 &   6.70~ &  0.41 &   8.74~ & 0.52 \\
F10        & --    ~ & --    & --    ~ & --    & --     & --   & --     & --    & --    ~ & --    & --    ~ & --   \\
F11        & --    ~ & --    &   3.49~ &  0.38 &   2.66 & 0.11 &   3.82 &  0.19 &   2.79~ &  0.17 &   1.77~ & 0.11 \\
F12        & --    ~ & --    &   3.47~ &  0.38 &   1.87 & 0.08 &   2.49 &  0.12 &   1.44~ &  0.09 &   1.15~ & 0.07 \\
F13        &  16.28: &  7.65 &   8.69~ &  0.81 &   8.71 & 0.36 &  14.17 &  0.71 &  12.30~ &  0.75 &  14.17~ & 0.85 \\
F14        & --    ~ & --    &   2.28: &  0.30 &   0.97 & 0.04 &   0.84 &  0.05 &   0.64~ &  0.04 &   0.41: & 0.03 \\
\tableline \tableline
\sidehead{RSGC2}
\tableline
St2-01     & --    ~ & --    & --    ~ & --    &   2.03 & 0.08 &   1.93 &  0.11 &   1.21~ &  0.08 & --    ~ & --   \\
St2-02     & --    ~ & --    & --    ~ & --    &   1.63 & 0.07 &   1.48 &  0.09 &   0.67: &  0.06 & --    ~ & --   \\
St2-03     & --    ~ & --    &  15.74: &  2.27 &  13.14 & 0.54 &  19.91 &  1.00 &  15.38~ &  0.94 &  13.19~ & 0.79 \\
St2-04     &  18.16: &  9.03 &  10.07: &  1.54 &   5.12 & 0.21 &   6.27 &  0.31 &   4.02~ &  0.25 &   2.49~ & 0.17 \\
St2-05     & --    ~ & --    & --    ~ & --    &   1.71 & 0.07 &   1.52 &  0.10 &   1.08~ &  0.08 & --    ~ & --   \\
\tableline \tableline
\sidehead{RSGC2 SW}
\tableline
St2-08     & --    ~ & --    & --    ~ & --    &   1.38 & 0.06 &   1.54 &  0.10 &   1.01~ &  0.07 & --    ~ & --   \\
St2-11     & --    ~ & --    & --    ~ & --    &   7.35 & 0.30 &  16.61 &  0.83 &  16.56~ &  1.01 &  14.25~ & 0.86 \\
St2-14     & --    ~ & --    & --    ~ & --    &   3.38 & 0.14 &   4.88 &  0.25 &   3.92~ &  0.24 &   2.94~ & 0.19 \\
St2-18     &  27.77~ &  2.83 &  48.62~ &  5.01 &  31.24 & 1.28 &  56.86 &  2.84 &  45.89~ &  2.80 &  35.74~ & 2.14 \\
St2-26     & --    ~ & --    & --    ~ & --    &   1.37 & 0.06 &   1.83 &  0.11 &   1.34~ &  0.09 & --    ~ & --   \\
\tableline \tableline 
\sidehead{Per OB1}
\tableline
W Per      &  50.99~ &  5.15 &  63.11~ &  5.87 &  63.45 & 2.60 &  79.47 &  3.97 &  62.61~ &  3.82 &  68.76~ &  4.13 \\
S Per      & 147.27~ & 12.81 & 186.16~ & 16.57 & 162.82 & 6.68 & 274.56 & 13.73 & 185.62~ & 11.32 & 178.67~ & 10.72 \\
T Per      &  29.18: &  3.73 &  19.93~ &  2.13 &  10.57 & 0.43 &   8.66 &  0.43 &   6.57~ &  0.40 &   4.75~ &  0.30 \\
V605 Cas   &  16.99: &  2.46 &  27.32~ &  2.71 &  12.54 & 0.51 &  11.32 &  0.57 &   7.69~ &  0.47 &   7.93~ &  0.48 \\
V778 Cas   & --    ~ & --    &  32.23~ &  3.35 &  10.22 & 0.42 &   7.20 &  0.38 &   4.94~ &  0.32 &   2.89: &  0.51 \\
PR Per     &  26.75: &  3.56 &  28.78: &  3.28 &  14.36 & 0.59 &  11.78 &  0.59 &   8.90~ &  0.54 &   8.26~ &  0.51 \\
FZ Per     & --    ~ & --    & --    ~ & --    &  12.19 & 0.50 &  10.90 &  0.54 &   6.52~ &  0.40 &   4.13~ &  0.27 \\
BD+59 372  & --    ~ & --    & --    ~ & --    &   2.80 & 0.11 &   1.57 &  0.10 &   0.89: &  0.07 & --    ~ & --    \\
XX Per     & --    ~ & --    & --    ~ & --    & --     & --   & --     & --    & --    ~ & --    & --    ~ & --    \\
HD 236947  & --    ~ & --    & --    ~ & --    & --     & --   & --     & --    & --    ~ & --    & --    ~ & --    \\
KK Per     & --    ~ & --    & --    ~ & --    & --     & --   & --     & --    & --    ~ & --    & --    ~ & --    \\
V550 Per   & --    ~ & --    &  16.67: &  2.32 &   5.19 & 0.21 &   3.51 &  0.22 &   1.81~ &  0.15 & --    ~ & --    \\
PP Per     &  32.92: &  9.81 &  12.56: &  2.10 &   5.48 & 0.22 &   2.74 &  0.17 &   2.31~ &  0.16 & --    ~ & --    \\
BU Per     &  33.14~ &  3.38 &  34.91~ &  3.56 &  33.10 & 1.36 &  36.70 &  1.83 &  26.30~ &  1.60 &  30.21~ &  1.81 \\
AD Per     & --    ~ & --    &  35.80~ &  3.37 &  20.75 & 0.85 &  20.44 &  1.02 &  13.12~ &  0.80 &  11.25~ &  0.69 \\
SU Per     & --    ~ & --    &  20.58~ &  2.41 &  43.77 & 1.79 &  40.03 &  2.00 &  24.11~ &  1.47 &  30.08~ &  1.80 \\
RS Per     & --    ~ & --    &  73.08~ &  6.72 &  57.59 & 2.36 &  59.47 &  2.97 &  42.78~ &  2.61 &  41.09~ &  2.47 \\
V439 Per   & --    ~ & --    &  17.79: &  2.38 &   7.33 & 0.30 &   5.18 &  0.27 &   3.20~ &  0.20 &   2.39~ &  0.17 \\
V403 Per   &  30.91~ &  3.21 &  15.85: &  2.25 &   5.31 & 0.22 &   2.71 &  0.16 &   1.83~ &  0.12 & --    ~ & --    \\
V441 Per   &  39.42~ &  3.86 &  44.46~ &  4.31 &  19.32 & 0.79 &  16.81 &  0.84 &  13.26~ &  0.81 &  11.10~ &  0.68 \\
YZ Per     &  37.91~ &  4.25 &  42.70~ &  4.18 &  33.62 & 1.38 &  31.78 &  1.59 &  21.23~ &  1.29 &  20.18~ &  1.21 \\
GP Cas     &  45.68~ &  4.29 &  48.57~ &  4.57 &  22.97 & 0.94 &  23.51 &  1.18 &  17.14~ &  1.05 &  16.10~ &  0.97 \\
V648 Cas   &  54.72~ &  5.03 &  50.40~ &  4.74 &  31.87 & 1.31 &  34.81 &  1.74 &  26.89~ &  1.64 &  24.57~ &  1.47 \\
\tableline \tableline
\sidehead{Mc8}
\tableline
Mc8-01     &   3.80: &  0.48 & --    ~ & --    &   4.66 & 0.19 &   5.61 &  0.28 &   4.68~ &  0.29 &   3.67~ & 0.22 \\
Mc8-02     &   2.82: &  0.42 & --    ~ & --    &   1.28 & 0.05 &   1.04 &  0.05 &   0.79~ &  0.05 &   0.73~ & 0.05 \\
Mc8-03     &  10.78: &  2.12 & --    ~ & --    &   3.39 & 0.14 &   4.12 &  0.21 &   3.12~ &  0.19 &   3.24~ & 0.19 \\
Mc8-04     &   3.23: &  0.45 & --    ~ & --    &   3.21 & 0.13 &   3.64 &  0.18 &   2.80~ &  0.17 &   2.04~ & 0.12 \\
Mc8-05     & --    ~ & --    & --    ~ & --    &   1.44 & 0.06 &   2.17 &  0.11 &   1.63~ &  0.10 &   1.02~ & 0.07 \\
Mc8-06     &   5.00~ &  0.55 & --    ~ & --    &   2.46 & 0.10 &   2.27 &  0.11 &   1.45~ &  0.09 &   0.75~ & 0.05 \\
\enddata

\tablenotetext{:}{The colon represents uncertain detection.}

\end{deluxetable}

\clearpage


\clearpage

\begin{deluxetable}{lrrrr}
\tabletypesize{\scriptsize}
\tablecaption{Infrared Photometric Data 5\label{tbl-11}}
\tablecolumns{5}
\tablewidth{0pt}
\tablehead{\colhead{Source} & \multicolumn{4}{c}{AKARI}                                                       \\ 
\cline{2-5}
\colhead{name}   & \colhead{9.0~$\mu$m} & \colhead{$\sigma_{9\micron}$} & \colhead{18.0~$\mu$m} & \colhead{$\sigma_{18\micron}$} \\
\colhead{}       & \colhead{(Jy)}       & \colhead{(Jy)}                & \colhead{(Jy)}        & \colhead{(Jy)}                 }
\startdata
\sidehead{Wd1}
\tableline 
W 237      &  91.65 & 7.35 & 113.80 & 0.67 \\
W 26       & 296.50 & 7.68 & 449.70 & 4.31 \\
W 20       &  61.41 & 0.28 & 111.70 & 0.67 \\
W 75       &  24.26 & 1.64 & --     & --   \\
\tableline \tableline
\sidehead{RSGC1}
\tableline
F01        & 11.47 & 0.34 & -- & --   \\
F02        & 12.69 & 0.17 & -- & --   \\
F03        &  6.46 & 0.54 & -- & --   \\
F04        &  5.75 & 0.03 & -- & --   \\
F05        &  7.86 & 0.45 & -- & --   \\ 
F06        &  5.17 & 0.58 & -- & --   \\
F07        & --    & --   & -- & --   \\
F08        &  7.86 & 0.45 & -- & --   \\
F09        & --    & --   & -- & --   \\
F10        & --    & --   & -- & --   \\
F11        &  4.01 & 0.70 & -- & --   \\
F12        &  2.16 & 0.12 & -- & --   \\
F13        & 13.30 & 0.41 & -- & --   \\
F14        &  0.91 & 0.01 & -- & --   \\
\tableline \tableline
\sidehead{RSGC2}
\tableline
St2-01     &  2.64 & 0.13 &  1.52 & 0.08 \\
St2-02     &  1.51 & 0.05 &  0.56 & 0.05 \\
St2-03     & 14.16 & 0.26 & 14.24 & 0.17 \\
St2-04     &  6.98 & 1.43 &  3.68 & 0.08 \\
St2-05     &  1.45 & 0.01 &  0.66 & 0.09 \\
\tableline \tableline
\sidehead{RSGC2 SW}
\tableline
St2-08     &  1.30 & 0.16 & --    & --   \\
St2-11     &  8.62 & 0.41 & 12.31 & 0.24 \\
St2-14     & --    & --   &  2.80 & 0.14 \\
St2-18     & 33.98 & 0.41 & 36.78 & 0.08 \\
St2-26     &  1.45 & 0.20 &  1.17 & 0.09 \\
\tableline \tableline
\sidehead{Per OB1}
\tableline
W Per      &  77.36 & 0.82 &  73.29 &  0.77 \\
S Per      & 349.70 & 3.68 & 285.50 & 11.80 \\
T Per      &   9.80 & 0.12 &   5.77 &  0.09 \\
V605 Cas   &  12.00 & 0.14 &   7.64 &  0.17 \\
V778 Cas   &   9.61 & 0.01 &   3.85 &  0.04 \\
PR Per     &  13.01 & 0.08 &   8.09 &  0.10 \\
FZ Per     &  12.03 & 0.12 &   4.98 &  0.05 \\
BD+59 372  &   2.52 & 0.02 &   0.66 &  0.02 \\
XX Per     &  64.90 & 1.57 &  28.47 &  0.40 \\
HD 236947  &   4.69 & 0.02 &   1.35 &  0.02 \\
KK Per     &  19.18 & 0.08 &   9.50 &  0.08 \\
V550 Per   &   4.63 & 0.03 &   1.67 &  0.01 \\
PP Per     &   4.76 & 0.05 &   1.43 &  0.01 \\
BU Per     &  36.59 & 0.51 &  30.27 &  0.51 \\
AD Per     &  20.81 & 0.15 &  11.90 &  0.13 \\
SU Per     &  45.48 & 0.47 &  30.69 &  0.41 \\
RS Per     &  66.55 & 0.88 &  47.17 &  0.85 \\
V439 Per   &   6.93 & 0.04 &   2.29 &  0.02 \\
V403 Per   &   4.81 & 0.02 &   1.33 &  0.03 \\
V441 Per   & --     & --   &  11.91 &  0.15 \\
YZ Per     &  37.24 & 0.36 &  25.05 &  0.08 \\
GP Cas     &  24.89 & 0.40 &  15.95 &  0.04 \\
V648 Cas   &  31.81 & 1.29 &  24.37 &  0.30 \\
\tableline \tableline
\sidehead{Mc8}
\tableline
Mc8-01     & 5.17 & 0.11 & --   & --   \\
Mc8-02     & 1.43 & 0.05 & 0.67 & 0.05 \\
Mc8-03     & --   & --   & 2.62 & 0.13 \\
Mc8-04     & --   & --   & 3.00 & 0.58 \\
Mc8-05     & --   & --   & 1.17 & 0.04 \\
Mc8-06     & --   & --   & --   & --   \\
\enddata

\end{deluxetable}

\clearpage


\begin{thebibliography}{}
\bibitem[Alexander et al.(2009)]{ale09} Alexander, M. J., Kobulnicky, H. A., Clemens, D. P., Jameson, K., Pinnick, A., \& Pavel, M. 2009, \apj, 137, 4824
\bibitem[An et al.(2007)]{and07} An, D., Terndrup, D. M., Pinsonneault, M. H., Paulson, D. B., Hanson, R. B., \& Stauffer, J. R. 2007, \apj, 655, 233
\bibitem[Asaki et al.(2010)]{asa10} Asaki, Y., Deguchi, S., Imai, H., Hachisuka, K., Miyoshi, M., \& Honma, M. 2010, \apj, 721, 267
\bibitem[Boboltz \& Claussen(2004)]{bob04} Boboltz, D. A., \& Claussen, M. J. 2004, \apj, 608, 480
\bibitem[Borgman et al.(1970)]{bor70} Borgman, J., Koornneef, J., \& Slingerland, J. 1970, \aap, 4, 248
\bibitem[Brandner et al.(2008)]{bra08} Brandner, W., Clark, J. S., Stolte, A., Waters, R., Negueruela, I., \& Goodwin, S. P.  2008, \aap, 478, 137
\bibitem[Chen \& Gao(2002)]{che02} Chen, P.-S., \& Gao, Y.-F. 2002, Chinese J. Astron. Astrophys., 2, 169
\bibitem[Cho et al.(1996)]{cho96} Cho, S.-H., Kaifu, N., \& Ukita, N. 1996, Astron. Astrophys. Suppl. Ser., 115, 117
\bibitem[Choi et al.(2008)]{choi08} Choi, Y. K., Hirota, T., Honma, M., Kobayashi, H., Bushimata, T., et al. 2008, \pasj, 60, 1007
\bibitem[Clark et al.(2005)]{cla05} Clark, J. S., Negueruela, I., Crowther, P. A., \& Goodwin, S. P. 2005, \aap, 434, 949
\bibitem[Clark et al.(2009)]{cla09} Clark, J. S., Negueruela, I., Davies, B., Larionov, V. M., Ritchie, B. W., Figer, D. F., Messineo, M., Crowther, P. A., \& Arkharov, A. A. 2009, \aap, 498, 109
\bibitem[Crowther et al.(2006)]{cro06} Crowther, P. A., Hadfield, L. J., Clark, J. S., Negueruela, I., Vacca, W. D., Krabbe, A., Genzel, R., Eckart, A., Najarro, F., Lutz, D., Cameron, M., Kroker, H., Tacconi-Garman, L. E., Thatte, N., Weitzel, L., Drapatz, S., Geballe, T., Sternberg, A., \& Kudritzki, R. 2006a, \mnras, 372, 1407
\bibitem[Crowther et al.(2008)]{cro08} Crowther, P. A., Hadfield, L. J., Clark, J. S., Negueruela, I., \& Vacca, W. D. 2008, \mnras, 385, 544
\bibitem[Davies et al.(2007)]{dav07} Davies, B., Figer, D. F., Kudritzki, R.-P. et al. 2007, \apj, 671, 781
\bibitem[Davies et al.(2008)]{dav08} Davies, B., Figer, D. F., Law, C. J. et al. 2008, \apj, 676, 1016
\bibitem[Deguchi et al.(1989)]{deg89} Deguchi, S., Nakada, Y., Forster, J. R. 1989, \mnras, 239, 825
\bibitem[Deguchi et al.(2010)]{deg10} Deguchi, S., Nakashima, J., Zhang, Y., et al. 2010, \pasj, 62, 391
\bibitem[Deul et al.(1995)]{deu95} Deul, E., Epchtein, N., Borsenberger, J. 1995, Information \& on-line data in astronomy, 203, 73
\bibitem[Dollery et al.(1987)]{dol87} Dollery, M. E., Gaylard, M. J., Cohen, R. J. 1987, IAUS, 122, 215
\bibitem[Egan et al.(1999)]{ega99} Egan M. P., Price S. D., Moshir, M. M., Cohen, M., \& Tedesco, E., 1999, The Mid-Course Space Experiment Point Source Catalog Version 1.2 Explanatory Guide (AFRLVS-TR-1999-1522)
\bibitem[Egan et al.(2003)]{ega03} Egan M.P., Price S.D., Kraemer K.E., Mizuno D.R., Carey S.J., Wright C.O., Engelke C.W., Cohen M., \& Gugliotti G. M. 2003, VizieR Online Data Catalog, 5114, 0
\bibitem[Figer et al.(2006)]{fig06} Figer, D. F., MacKenty, J. W., Robberto, M. et al. 2006, \apj, 643, 1166
\bibitem[Fouque et al.(1992)]{fou92} Fouque, P., Le Bertre, T., Epchtein, N., Guglielmo, F., Kerschbaum, F. 1992, \aaps, 93, 151
\bibitem[Haikala et al.(1994)]{hai94} Haikala, L. K., Nyman, L.-\AA., \& Forsstrom, V. 1994, \apss, 103, 107 
\bibitem[Humphreys(1970)]{hum70} Humphreys, R. M. 1970, \apj, 160, 1149
\bibitem[Henry \& Worthey(1999)]{hen99} Henry, R. B. C. \& Worthey, G. 1999, PASP, 111, 919
\bibitem[Indebetouw et al.(2005)]{ind05} Indebetouw, R., Mathis, J. S., Babler, B. L., Meade, M. R., Watson, C., Whitney, B. A., Wolff, M. J., Wolfire, M. G., Cohen, M., Bania, T. M., Benjamin, R. A., Clemens, D. P., Dickey, J. M., Jackson, J. M., Kobulnicky, H. A., Marston, A. P., Mercer, E. P., Stauffer, J. R., Stolovy, S. R., \& Churchwell, E. 2005, \apj, 619, 931 
\bibitem[Ishihara et al.(2010)]{ish10} Ishihara, D., Onaka, T., Kataza, H., Salama, A., Alfageme, C., Cassatella, A., Cox, N., Garc\'{i}a-Lario, P., Stephenson, C., Cohen, M., Fujishiro, N., Fujiwara, H., Hasegawa, S., Ita, Y., Kim, W., Matsuhara, H., Murakami, H., M\"{u}ller, T. G., Nakagawa, T., Ohyama, Y., Oyabu, S., Pyo, J., Sakon, I., Shibai, H., Takita, S., Tanab\'{e}, T., Uemizu, K., Ueno, M., Usui, F., Wada, T., Watarai, H., Yamamura, I., \& Yamauchi, C. 2010, \aap, 514, A1
\bibitem[Ivezi\'{c} et al.(1999)]{ive99} Ivezi\'{c}, Z., Nenkova, M., \& Elitzur, M. 1999, User Manual for DUSTY, University of Kentucky Internal Report
\bibitem[Jewell et al.(1984)]{jew84} Jewell, P. R., Batrla, W., Walmsley, C. M., \& Wilson, T. L. 1984, \aap, 130, L1
\bibitem[Jiang et al.(1996)]{jia96} Jiang, B. W., Deguchi, S., Yamamura, I., Nakada, Y., Cho, S. H., \& Yamagata, T. 1996, \apjs, 106, 463
\bibitem[Jiang et al.(1999)]{jia99} Jiang, B. W., Deguchi, S., \& Ramesh, B. 1999, \pasj, 51, 95
\bibitem[Josselin \& Plez(2005)]{jos05} Josselin, E., \& Plez, B. 2005, ESO Astrophysics Symposia, 2005, 405
\bibitem[Kamohara et al.(2005)]{kam05} Kamohara, R., Deguchi, S., Miyoshi, M., \& Shen, Z.-Q. 2005, \pasj, 57, 341
\bibitem[Kothes \& Dougherty(2007)]{kot07} Kothes, R., \& Dougherty, S. M. 2007, \aap, 468, 993
\bibitem[Le Sidaner \& Le Bertre(1996)]{sid96} Le Sidaner, P., \& Le Bertre, T. 1996, \aap, 314, 896
\bibitem[Levesque(2010)]{lev09} Levesque, E. M. 2010, \nar, 54, 1
\bibitem[Levesque et al.(2005)]{lev05} Levesque, E. M., Massey, P., Olsen, K. A., Plez, B., Josselin, E., Maeder, A., \& Meynet, G. 2005, \apj, 628, 973
\bibitem[Luna et al.(2009)]{lun09} Luna, A., Mayya, Y. D., Carrasco, L., Rodr\'{i}guez-Merino, L. H., \& Bronfman, L. 2009, RevMexAA, 37, 32
\bibitem[Mathis et al.(1977)]{mat77} Mathis, J. S., Rumpl, W. \& Nordsieck, K. H. 1977, \apj, 17, 25
\bibitem[Mauron \& Josselin(2011)]{mau11} Mauron, N., \& Josselin, E. 2011, \aap, 526, A156
\bibitem[Mengel \& Tacconi-Garman(2009)]{men09} Mengel, S., \& Tacconi-Garman, L. E. 2009, \apss, 324, 321
\bibitem[Mengel \& Tacconi-Garman(2007)]{men07} Mengel, S., \& Tacconi-Garman, L. E. 2007, \aap, 466, 151
\bibitem[Messineo et al.(2002)]{mes02} Messineo, M., Habing, H. J., Sjouwerman, L. O., Omont, A., \& Menten, K. M. 2002, \aap, 393, 115
\bibitem[Messineo et al.(2012)]{mes12} Messineo, M., Menten, K. M., Churchwell, E., \& Habing, H. 2012, \aap, 537, A10
\bibitem[Nakashima \& Deguchi(2006)]{nak06} Nakashima, J., \& Deguchi, S. 2006, \apjl, 647, L139
\bibitem[Nakashima \& Deguchi(2007)]{nak07} Nakashima, J., \& Deguchi, S. 2007, \apj, 669, 446
\bibitem[Negueruela et al.(2010a)]{neg10a} Negueruela, I., Gonzalez-Fernandez, C., Marco, A., Clark, J. S., \& Martinez-Nunez, S. 2010, \aap, 513, A74
\bibitem[Negueruela et al.(2010b)]{neg10b} Negueruela, I., Clark, J. S., \& Ritchie, B. W. 2010, \aap, 516, A78
\bibitem[Negueruela et al.(2011)]{neg11} Negueruela, I., Gonzalez-Fernandez, C., Marco, A., \& Clark, J. S. 2011, \aap, 528, A59
\bibitem[Nishiyama et al.(2006)]{nis06} Nishiyama, S., Nagata, T., Kusakabe, N., Matsunaga, N., Naoi, T., et al. 2006, \apj, 638, 839
\bibitem[Ossenkopf et al.(1992)]{oss92} Ossenkopf, V., Henning, Th., \& Mathis, J. S. 1992, \aap, 261, 567
\bibitem[Palagi et al.(1993)]{pal93} Palagi, F., Cesaroni, R., Comoretto, G., Felli, M., \& Natale, V. 1993, \aaps, 101, 153
\bibitem[Pashchenko et al.(2006)]{pas06} Pashchenko, M. I., Rudnitskij, G. M., Samodurov, V. A., \& Tolmachev, A. M. 2006, Astron. Astrophys. Trans., 25, 399
\bibitem[Perryman et al.(1998)]{per98} Perryman, M. A. C., Brown, A. G. A., Lebreton, Y., G\'{o}mez, A., Turon, C., Cayrel de Strobel, G., Mermilliod, J. C., Robichon, N., Kovalevsky, J., \& Crifo, F. 1998, \aap, 331, 81
\bibitem[Reid et al.(2009)]{rei09} Reid, M. J., Menten, K. M., Zheng, X. W., Brunthaler, A., Moscadelli, L., Xu, Y., Zhang, B., Sato, M., Honma, M., Hirota, T., Hachisuka, K., Choi, Y. K., Moellenbrock, G. A., \& Bartkiewicz, A. 2009, \apj, 700, 137
\bibitem[Sault et al.(1995)]{sau95} Sault, R. J., Teuben, P. J., \& Wright, M. C. H. 1995, Astronomical Data Analysis Software and Systems IV, ASP Conference Series, eds, Shaw, R. A., Payne, H. E., \& Hayes, J. J. E., 77, 433
\bibitem[Scholz \& Wood(2000)]{sch00} Scholz, M., \& Wood, P. R. 2000, \aap, 362, 1065
\bibitem[Skrutskie et al.(2006)]{skr06} Skrutskie, M. F., Cutri, R. M., Stiening, R., Weinberg, M. D., Schneider, S., Carpenter, J. M., Beichman, C., Capps, R., Chester, T., Elias, J., Huchra, J., Liebert, J., Lonsdale, C., Monet, D. G., Price, S., Seitzer, P., Jarrett, T., Kirkpatrick, J. D., Gizis, J. E., Howard, E., Evans, T., Fowler, J., Fullmer, L., Hurt, R., Light, R., Kopan, E. L., Marsh, K. A., McCallon, H. L., Tam, R., Van Dyk, S., \& Wheelock, S. 2006, \aj, 131, 1163 
\bibitem[Stothers \& Chin(1999)]{sto99} Stothers, R. B., \& Chin, C-W. 1999, \apj, 522, 960
\bibitem[Su et al.(2012)]{su12} Su, J. B., Shen, Z.-Q., Chen, X., Yi, Jiyune, Jiang, D. R., \& Yun, Y. J. 2012, \apj, 754, 47
\bibitem[Szymczak et al.(2010)]{szy10} Szymczak, M., Woolak, P., G\'{e}rard, E., \& Richards, A. M. S. 2010, \aap, 524, A99
\bibitem[Takaba et al.(2001)]{tak01} Takaba, H., Iwata, T., Miyaji, T., \& Deguchi, S. 2001, \pasj, 53, 517
\bibitem[Verheyen et al.(2012)]{ver12} Verheyen, L., Messineo, M., \& Menten, K. M. 2012, \aap, 541, A36
\bibitem[Werner et al.(2004)]{wer04} Werner, M. W., Roellig, T. L., Low, F. J., Rieke, G. H., Rieke, M., Hoffmann, W. F., Young, E., Houck, J. R., Brandl, B., Fazio, G. G., Hora, J. L., Gehrz, R. D., Helou, G., Soifer, B. T., Stauffer, J., Keene, J., Eisenhardt, P., Gallagher, D., Gautier, T. N., Irace, W., Lawrence, C. R., Simmons, L., Van Cleve, J. E., Jura, M., Wright, E. L., \& Cruikshank, D. P. 2004, \apjs, 154, 1
\bibitem[Wilson \& Barrett(1972)]{wil72} Wilson, W. J., \& Barrett, A. H. 1972, \aap, 17, 385
\bibitem[Wright et al.(2010)]{wri10} Wright, Edward L.; Eisenhardt, Peter R. M., Mainzer, A. K., Ressler, M. E., Cutri, R. M., Jarrett, T., Kirkpatrick, J. D., Padgett, D., McMillan, R. S., Skrutskie, M., Stanford, S. A., Cohen, M., Walker, R. G., Mather, J. C., Leisawitz, D., Gautier, T. N., III, McLean, I., Benford, D., Lonsdale, C. J., Blain, A., Mendez, B., Irace, W. R., Duval, V., Liu, F., Royer, D., Heinrichsen, I., Howard, J., Shannon, M., Kendall, M., Walsh, A. L., Larsen, M., Cardon, J. G., Schick, S., Schwalm, M., Abid, M., Fabinsky, B., Naes, L., \& Tsai, C-W. 2010, \aj, 140, 1868
\bibitem[Zhang et al.(2012a)]{zha12a} Zhang, B., Reid, M. J., Menten, K. M., \& Zheng, X. W. 2012a, \apj, 744, 23
\bibitem[Zhang et al.(2012b)]{zha12b} Zhang, B., Reid, M. J., Menten, K. M., Zheng, X. W., \& Brunthaler, A. 2012b, \aap, 544, A42

\end{thebibliography}
\end{document}